\documentclass[]{pasj01}
\draft

\Received{}
\Accepted{}
 
\begin{document} 

\title{ 
Time-resolved spectroscopy and photometry of an M dwarf flare star
YZ Canis Minoris with OISTER and TESS: Blue asymmetry in H$\alpha$
line during the non-white light flare
}

\author{Hiroyuki \textsc{Maehara}\altaffilmark{1}}
\altaffiltext{1}{Subaru Telescope Okayama Branch Office, National Astronomical Observatory of Japan,
 National Institutes of Natural Sciences, 
 3037-5 Honjo, Kamogata, Asakuchi,
Okayama 719-0232,
Japan}
\email{hiroyuki.maehara@nao.ac.jp}

\author{Yuta \textsc{Notsu}\altaffilmark{2,3}}
\altaffiltext{2}{Laboratory for Atmospheric and Space Physics, University of Colorado Boulder, 3665 Discovery Drive, Boulder, CO 80303, USA}
\altaffiltext{3}{National Solar Observatory, University of Colorado Boulder, 3665 Discovery Drive, Boulder, CO 80303, USA.}


\author{Kousuke \textsc{Namekata}\altaffilmark{4}}
\altaffiltext{4}{Department of Astronomy, Kyoto University, Kitashirakawa-Oiwake-cho, Sakyo, Kyoto
606-8502, Japan}

\author{Satoshi \textsc{Honda}\altaffilmark{5}}
\altaffiltext{5}{Nishi-Harima Astronomical Observatory, Center for Astronomy, University of Hyogo, Sayo,
Sayo, Hyogo 679-5313, Japan}

\author{Adam F. \textsc{Kowalski}\altaffilmark{2,3,6}}
\altaffiltext{6}{Department of Astrophysical and Planetary Sciences, University of Colorado Boulder, 2000 Colorado Ave, Boulder, CO 80305, USA.}

\author{Noriyuki \textsc{Katoh}\altaffilmark{5,7}}
\author{Tomohito \textsc{Ohshima}\altaffilmark{5}}
\altaffiltext{7}{Graduate School of Human Development and Environment, Kobe University, 3-11 Tsurukabuto, Nada-ku, Kobe, Hyogo 657-8501, Japan}

\author{Kota \textsc{Iida}\altaffilmark{8}}
\altaffiltext{8}{Department of Physics, Tokyo Institute of Technology, 2-12-1 Ookayama, Meguro-ku, Tokyo
152-8551, Japan}
\author{Motoki \textsc{Oeda}\altaffilmark{8}}
\author{Katsuhiro L. \textsc{Murata}\altaffilmark{8}}

\author{Masayuki \textsc{Yamanaka}\altaffilmark{9,10}} 
\altaffiltext{9}{Hiroshima Astrophysical Science Center, Hiroshima University,
Higashi-Hiroshima, Hiroshima 739-8526, Japan
}
\altaffiltext{10}{Okayama Observatory, Kyoto University, 3037-5 Honjo, Kamogata,
Asakuchi, Okayama 719-0232, Japan
}
\author{Kengo \textsc{Takagi}\altaffilmark{11}}
\altaffiltext{11}{Department of Physical Science, Hiroshima University, Kagamiyama
1-3-1, Higashi-Hiroshima 739-8526, Japan
}
\author{Mahito \textsc{Sasada}\altaffilmark{9}} 
\author{Hiroshi \textsc{Akitaya}\altaffilmark{9}}

\author{Kai \textsc{Ikuta}\altaffilmark{4}}
\author{Soshi \textsc{Okamoto}\altaffilmark{4}}

\author{Daisaku \textsc{Nogami}\altaffilmark{4}}
\author{Kazunari \textsc{Shibata}\altaffilmark{12}}
\altaffiltext{12}{Kwasan Observatory, Yamashina, Kyoto 607-8471, Japan
}


\maketitle

\begin{abstract}
In this paper, we present the results from spectroscopic
and photometric observations of the M-type flare star 
YZ CMi in the framework of the Optical and Infrared 
Synergetic Telescopes for Education and Research (OISTER)
collaborations
during the Transiting Exoplanet Survey Satellite (TESS)
observation period.
We detected 145 white-light flares from the TESS light curve
and 4 H$\alpha$ flares from the OISTER observations
performed between 2019-01-16 and 2019-01-18.
Among them, 3 H$\alpha$ flares were associated with white-light flares.
However, one of them did not show clear brightening in continuum;
during this flare,  the H$\alpha$ line exhibited
blue-asymmetry which has lasted for $\sim 60$ min. 
The line of sight velocity of the blue-shifted
component is $-80$ -- $-100$ km s$^{-1}$.
This suggests that there can be upward flows of chromospheric cool
plasma 
even without detectable red/NIR continuum brightening.
By assuming that the blue-asymmetry in H$\alpha$ line was caused by
a prominence eruption on YZ CMi, 
we estimated the mass and kinetic energy of the upward-moving material
to be $10^{16}$ -- $10^{18}$ g and $10^{29.5}$ -- $10^{31.5}$ erg, respectively.
The estimated mass is comparable to expectations 
from the empirical relation between the flare X-ray 
energy and mass of upward-moving material for stellar flares and solar CMEs.
In contrast, the estimated kinetic energy for the non-white-light 
flare on YZ CMi is roughly $2$ orders of magnitude smaller than that
expected from the relation between flare X-ray energy and 
kinetic energy for solar CMEs.
This could be understood by the difference in the velocity between
CMEs and prominence eruptions.
\end{abstract}

\section{Introduction}
Solar flares are sudden and energetic explosions in the solar
atmosphere around sunspots. 
Flares are observed in all wavelength bands from radio to 
high-energy gamma-rays.
They are thought to be the rapid releases of magnetic energy 
through magnetic reconnection in the solar corona (e.g. \cite{ShibataMagara2011} and referenes theirin).
Part of the energy released by the magnetic reconnection
is transported from the reconnection site into the chromosphere
via thermal conduction and high-energy particles, 
which causes heating, produces line emission (e.g., H$\alpha$), and can
even produce hard X-ray and optical continuum emission.
This intense heating of chromospheric plasma is thought to cause
the upward flow of plasma called chromospheric evaporation \citep{Fisher1985}.

Similar periods of rapid increases and slow decays of intensity
in radio, optical and X-ray bands
are also observed on various types of stars, 
and they are called stellar flares.
In particular, it is known that
young stellar objects, close binary systems 
and M-type main sequence stars (dMe stars) exhibit 
frequent and energetic flares 
(e.g., \cite{ShibataYokoyama2002}; \cite{Gershberg2005}; \cite{Reid2005}; \cite{Benz2010}; \cite{Hawley2014}; \cite{Linsky2019}; \cite{Namekata2020}).
Because of the similarity in observational properties between
stellar flares and solar flares 
(e.g., Nuepert effect in solar/stellar flares; \cite{Neupert1968,Dennis1993,Hawley1995,Guedel1996}), they are considered
to be caused by the same physical processes (i.e., 
plasma heating by accelerated particles and evaporation).
Many spectroscopic studies of superflares have been
carried out in order to understand the dynamics of plasma during
flares and radiation mechanisms of flares.
Various spectroscopic observations of solar flares have shown that 
chromospheric lines (e.g., H$\alpha$, Ca II, Mg II) often exhibit 
asymmetric line profiles during flares.
Red asymmetries (enhancement of the red wing) have been frequently observed
during the impulsive phase of the flares 
(e.g. \cite{Svestka1962,Ichimoto1984,Canfield1990,Shoji1995,Berlicki2007,Kuridze2015,Kowalski2017,Graham2020}).
This is thought to be
caused by the chromospheric condensation, which is the downward flow of 
cool plasma in the chromosphere. 
Blue asymmetries (enhancement of the blue wing) 
have also been observed mainly in the early phase of flares
(e.g., \cite{Svestka1962}; \cite{Canfield1990}; \cite{Heinzel1994_BlueAsym}; \cite{Kuridze2016}; \cite{Tei2018}; \cite{Huang2019}). 
It is suggested that blue asymmetry is caused by an upflow of cool plasma,
which is lifted up by expanding hot plasma owing to the deep penetration of non-thermal electrons into the chromosphere during a flare (\cite{Tei2018}; \cite{Huang2019}). 
However, the detailed origin of this blue asymmetries are
still controversial.

Similar line asymmetries in chromospheric lines (especially H$\alpha$) have been 
observed during stellar flares.
In addition to red asymmetries (e.g., \cite{Houdebine1993}),
various blue asymmetries have been widely observed 
(e.g., \cite{Houdebine1990}; \cite{Gunn1994}; \cite{Fuhrmeister2008};
\cite{Vida2016}; \cite{Honda2018}; \cite{Muheki2020}).
\cite{Vida2016} reported several H$\alpha$ flares on the M4 dwarf
V374 Peg showing blue asymmetries with the line-of-sight velocity
of $-200$ -- $-400$ km s$^{-1}$. They also found that
red-wing enhancements in the H$\alpha$ line were observed
after blue asymmetries, which suggest the erupted cool plasma falling back
on the stellar surface.
\cite{Honda2018} reported a long duration H$\alpha$ flare on the M4.5 dwarf EV Lac.
During this flare, a blue asymmetry in the H$\alpha$ line with the dopper velocity
of $\sim -100$ km s$^{-1}$ has been observed for $> 2$ hours.
Since we cannot obtain the spatially resolved
information for stellar flares, 
the line of sight motions of cool plasma such as coronal rains,
surges, and filament/prominence eruptions may also cause
red/blue asymmetries.
For example, if the cool plasma is launched upward and 
seen above the limb, the emission 
can cause blue- or red-shifted enhancements
of the H$\alpha$ line (e.g. \cite{Odert2020}).
As observed on the Sun, such eruptions (surges and filament/prominece eruptions) 
can evolve into CMEs (coronal mass ejections)
if the erupted plasma is accelerated and the velocity exceeds 
the escape velocity (e.g., \cite{Gopalswamy2003}, \cite{ShibataMagara2011}, and reference therein).

Other studies have suggested that blue asymmetries in chromospheric lines
may be due to stellar mass ejections.
\citet{Vida2019} reported a statistical analysis of 478 stellar events
with asymmetries in Balmer lines of M-dwarfs, which were found from more than 5500 ``snapshot" spectra (cf. Similar events were also reported from other snapshot data in \cite{Fuhrmeister2018}). The velocity
and mass of the possible ejected materials estimated from the
blue- or red-shifted excess in Balmer lines range
from $100$ -- $300$ km s$^{-1}$ and $10^{15}$ -- $10^{18}$ g, respectively.
\citet{Moschou2019} presented the correlations between mass/kinetic energy
of CMEs and X-ray energy of associated flares on various types of stars.
They found that estimated stellar flare CME masses are consistent 
with the trends extrapolated from solar events but kinematic 
energies are roughly two orders of magnitude smaller than expected.
It is important to understand the properties of stellar CMEs 
in order to evaluate effects of stellar activities
not only on the mass- and angular momentum loss of the star (e.g. \cite{Osten2015}; \cite{Odert2017}, \cite{Cranmer2017}),
but also on the habitability (e.g., loss of atmosphere, atmospheric chemistry, climate, radiation dose) of exoplanets (e.g., \cite{Lammer2007}, \cite{Linsky2019}, \cite{Segura2010}, \cite{Tilley2019}, \cite{Scheucher2018}, \cite{Airapetian2020},  \cite{Yamashiki2019}).
However, our understanding of asymmetries in chromospheric lines 
and their connections with stellar flares/CMEs is still limited by the low
number of samples observed in time-resolved spectroscopy
simultaneously with high-precision photometry. 

In order to investigate the connection between the 
blue/red asymmetries in H$\alpha$ line
and properties of flares, we conducted photometric and spectroscopic
observations of an active M dwarf, YZ CMi, during
the Transiting Exoplanet Survey Satellite (TESS; \cite{Ricker2015})
observation window (sector 7: from 2019-01-07 to 2019-02-01).
In this paper, we report on results from  
the statistical analysis of flares on YZ CMi from TESS light curve
and photometric and spectroscopic observations 
in the framework of the Optical and Infrared 
Synergetic Telescopes for Education and Research 
(OISTER; Yamanaka et al. in prep.).
The details of our observations and analysis are
described in section 2.
We present the properties of detected flares from TESS and 
OISTER observations in section 3.
In section 4, we discuss (1) the differences between
H$\alpha$ flares with and without white-light flares,
(2) blue asymmetry observed during the
H$\alpha$ flare without a white-light flare,
(3) the rotational modulations observed in continuum and H$\alpha$ emission
line, and (4) the statistical properties of flare duration.

\section{Data and methods}
\subsection{Target star: YZ CMi}
YZ CMi (= Gl 285 = Ross 882) is a well-known 11 mag 
M4.5Ve flare star, whose distance from the Earth is about 5.99 pc \citep{Gaia2018}.
Flares on YZ CMi were first discovered at optical wavelengths
by \citet{vanMaanen1945}
and were later detected at radio wavelengths \citep{Lovell1969,Spangler1974}
and X-ray wavelengths \citep{Grindlay1975}. 
Frequent stellar flares have been observed on YZ CMi in several wavelength ranges
\citep{Lacy1976,Mitra-Kraev2005,Kowalski2013}, 
and in particular, a large superflare whose U-band energy is 
larger than 10$^{34}$ erg is reported in \citet{Kowalski2010}. 
Zeeman broadening measurements suggest the existence of strong magnetic 
fields on the stellar surface (e.g., \cite{Johns-Krull2000, Reiners2007}).
According to Zeeman–Doppler Imaging observation, 
the visible pole of YZ CMi is covered by a strong spot with the
radial magnetic field strength of up to $3$ kG \citep{Morin2008}.

\subsection{Flare detection from the TESS light curve}
We analyzed the TESS Sector 7 PDC-SAP light curve \citep{Vanderspek2018,Fausnaugh2019} 
of YZ CMi retrieved from
the MAST Portal site\footnote{https://mast.stsci.edu/portal/Mashup/Clients/Mast/Portal.html}.
In order to detect small flares, we first removed non-flare signals
such as long-term trend and rotational brightness variations from the light curve
and then searched for flares.
Since there is a data gap between BJD 2458503.04 and 2458504.71 in the TESS
light curve (Fig. \ref{fig:TESS_LC1}), 
we divided the whole TESS light curve into two subsets
(BJD 2458491.64 -- 2458503.04 and BJD 2458504.71 -- 2458516.09), and
analyzed each light curve subset separately.
First we removed some of the large flares from the light curve beforehand 
since large flares with long flare duration affect the long-term light 
curve fitting process.
The initial flare detection was performed
with the same method used in our previous studies \citep{Maehara2012,Shibayama2013}.
The excluded data points in the light curve were interpolated by using ``interpolate.Akima1DInterpolator'' in the Python Scipy package.
After removing flares and interpolation of the removed data points, 
we extracted the long-term trend and rotational variations
by using the 5th-order Bessel filter in the SciPy signal module.
We used the cut-off frequency of $0.2\ \rm d^{-1}$ in this process.
Then we removed the extracted signal from light curves.
We selected the data points satisfying the following conditions as the flare candidates:
(1) the residual brightness of the data point is higher than 
the upper 10 percent of the residual light curve, 
(2) at least two consecutive data points exceed the flare detection threshold,
(3) decay time is longer than rise time.
Finally we checked all the light curves of flare candidates by eye
and eliminated the misidentified candidates.
The total number of the automatically selected flare candidates is 194
and the number of confirmed flares is 145.
A table of the flares identified in TESS (with peak time, 
flare amplitude normalized by the average stellar brightness, 
equivalent duration (the model-independent energy in units of seconds; \cite{Gershberg1972}), bolometric energy of flare,
and flare e-folding time) is provided in Supplementary Data (Table\_S1.csv).

\subsection{OISTER observations}
 We conducted the coordinated observing campaign of YZ CMi on 2019-01-16, 17, and 18
 in the framework of OISETR collaboration.
 The log of the observation is summarised in Table \ref{OISTER_OBS}.
 
 Simultanous multi-color ($g$-, $R_C$-, and $I_C$-bands \footnote{
The effective wavelength ($\lambda _0$) and bandwidth
(FWHM)
are $\lambda _0=4770$ \AA\ and FWHM$=1380$ \AA\ 
for $g$-band \citep{Fukugita1996},
$\lambda _0=6407$ \AA\ and FWHM$=1580$ \AA\ 
for $R_C$-band \citep{Bessell2005},
$\lambda _0=7980$ \AA\ and FWHM$=1540$ \AA\ 
for $I_C$-band \citep{Bessell2005}, respectively.
}) photometry 
 was carried out by using the MITSuME 50-cm telescope at Okayama, Japan \citep{Kotani2005}.
 All the images taken by MITSuME were dark subtracted and flat-fielded
 by using IRAF \footnote{IRAF is distributed by the National Optical Astronomy Observatories, which are operated by the Association of Universities for Research in Astronomy, Inc., under cooperate agreement with the National Science Foundation.} 
 in the standard manner before the photometry was performed.
 We carried out aperture photometry of YZ CMi and several surrounding stars
 on each image with the APPHOT package in IRAF.
 We used a nearby K III star HD 62525 ($V=8.072, B-V=0.810$; magnitude and color are 
 taken from the AAVSO Variable Star Plotter (VSP)\footnote{https://www.aavso.org/apps/vsp/}) 
 as a local standard star for photometry.
 The constancy of brightness of the comparison star during our observation
 was checked by using TYC 183-2106-1 ($V=10.661, B-V=0.458$; taken from VSP).
 
 We also performed time-resolved low ($R=\lambda/\Delta \lambda \sim 400$)
 and medium ($R=\lambda/\Delta \lambda \sim 10000$) resolution spectroscopy
 by using the HOWPol \citep{Kawabata2008} mounted on the 1.5-m Kanata telescope 
 at Higashi Hiroshima Observatory, Hiroshima University
 and the MALLS
 (Medium And Low-dispersion Long-slit Spectrograph; \cite{Ozaki2005})
 mounted on the 2.0-m Nayuta telescope
 at Nishi-Harima Astronomical Observatory, University of Hyogo, respectively.
 After the standard image reduction processes such as dark-subtraction and
 flat-fielding, we analyzed the data by using the TWODSPEC and ONEDSPEC
 packages in IRAF.
 For the spectra taken with Kanata/HOWPol, the wavelength calibration
 was performed by using O\,I and Hg sky glow lines.
 For the spectra taken with Nayuta/MALLS, we used an Fe/Ne/Ar lamp for
 wavelength calibration. In addition to the standard wavelength
 calibration procedure by using the comparison frames, we corrected
 the instrumental drift of the spectrum in the wavelength dimension over time 
 by using the atmospheric absorption features. 
 We also applied corrections for the barycentric velocity
 and the absolute stellar radial velocity (26.495 km s$^{-1}$; \cite{Soubiran2018})
 to the wavelength of spectra.
\subsection{APO observations}
In this paper, we used Apache Point Observatory (APO) data 
only for rotational modulations 
discussed in Section 4.3. 
The detailed results including flares from 
the APO observations will be presented in our forthcoming paper 
(Notsu et al. in prep.).
The log of the observation is also summarised in Table \ref{OISTER_OBS}.
We carried out $g$-band photometry with the Flarecam instrument 
\citep{Hilton2011} of 0.5-m Astrophysical Research Consortium Small 
Aperture Telescope (ARCSAT) at APO 
on 2019-01-26, 27, and 28 (UT). 
Dark subtraction and flat-fielding were performed by using PyRAF \footnote{
PyRAF is part of the stscipython package of astronomical data analysis tools and is a product of the Science Software Branch at the Space Telescope Science Institute.
}
software in the standard manner before the photometry. 
Aperture photometry was performed by using AstroimageJ 
\footnote{https://www.astro.louisville.edu/software/astroimagej/}
\citep{Collins2017}. 
We used nearby stars as the magnitude reference. 
Spectroscopic observations were carried out by using 
the ARC Echelle Spectrograph (ARCES; \cite{Wang2003}) 
attached to the ARC 3.5 m telescope at APO on 2019-01-26, 27, and 28 (UT). 
The wavelength resolution ($R=\lambda/\Delta \lambda$) 
is $\sim$ 32000, and the spectral coverage is $3800$–-$10000$ \AA
. After the standard image reduction procedures such as bias subtraction, 
flat-fielding, and scattered light subtraction, we analyzed the data 
by using the ECHELLE package in IRAF and PyRAF software. 
We used a Th/Ar lamp for wavelength calibration. We also applied the 
heliocentric radial velocity correction by using the ECHELLE package. 
These analysis methods of spectroscopic data are the same as 
\citet{Notsu2019}. The H$\alpha$ equivalent width values 
are measured from these spectra.



\begin{table}
  \tbl{Log of OISTER and APO observations. }
  {\begin{tabular}{crcc}
    \hline
Telescope/Instrument & Start-End (UT) & Exp. time (sec) & Number of data\\
 \hline
MITSuME 0.5m & 2019-01-16.625 - 16.793 & 5 &  590 ($g$), 608 ($R_C$), 610 ($I_C$) \\
($g,R_C,I_C$-band) &         17.713 - 17.786 & 5 & 123 ($g$), 124 ($R_C$), 124 ($I_C$)\\ 
 Okayama, Japan       &         18.623 - 18.793 & 5 & 853 ($g$), 891 ($R_C$), 895 ($I_C$)\\\hline
Kanata 1.5m/HOWPol &         16.625 - 16.816 & 60 &  224\\
($4000$-$9000$\AA;$\lambda/\Delta \lambda \sim 400$) &         17.667 - 17.740 & 60 & 77\\
 Hiroshima, Japan  &         18.639 - 18.792 & 60 & 186\\\hline
Nayuta 2m/MALLS  &       16.604 - 16.801 & 250 & 54 \\
($6350$-$6800$\AA; $\lambda/\Delta \lambda \sim 10000$) &         18.629 - 18.793 & 250 & 51\\
 Hyogo, Japan & & & \\\hline
ARCSAT 0.5m/flarecam & 26.131 - 26.423  & 4, 15, 30\footnotemark[$*$] & 310 ($g$) \\
 ($g$-band)     & 27.113 - 27.418  & 4 & 138 ($g$) \\
 New Mexico, United States               & 28.108 - 28.402  & 4, 6, 12, 20\footnotemark[$*$] & 155 ($g$) \\\hline
ARC 3.5m/ARCES & 26.118 - 26.420 & 600, 900\footnotemark[$*$] & 28 \\
($3800$–$10000$\AA; $\lambda/\Delta \lambda \sim 32000$)         & 27.110 - 27.421 & 300 & 60 \\
 New Mexico, United States             & 28.112 - 28.413 & 300, 600\footnotemark[$*$] & 54 \\
         \hline
  \end{tabular}}\label{OISTER_OBS}
    \begin{tabnote}
 \footnotemark[$*$] We adjusted the exposure time as needed since the sky was covered by a thin layer of clouds.
  \end{tabnote}
\end{table}

\begin{figure}
 \begin{center}
  \includegraphics[width=16cm]{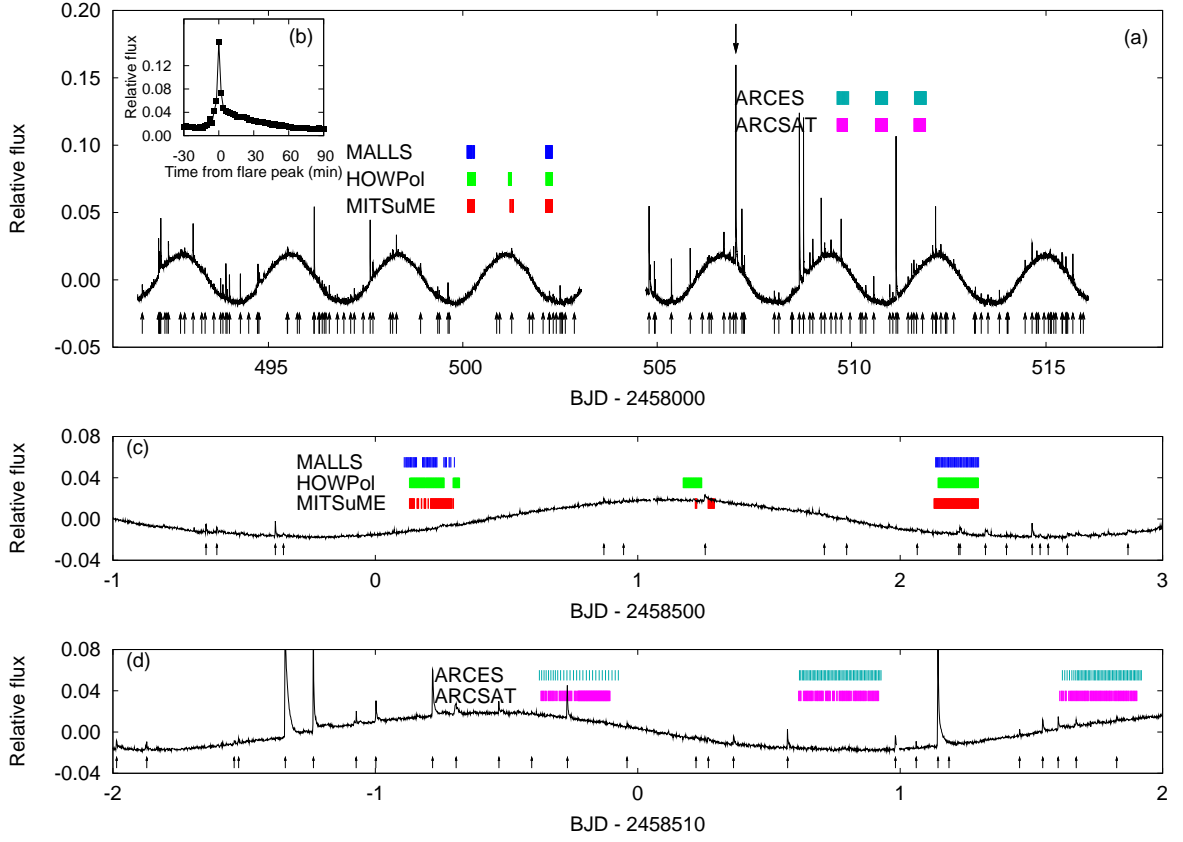} 
 \end{center}
\caption{
(a) Light curve of YZ CMi observed with TESS, 
covering the times of OISTER and APO observations. 
The horizontal and vertical axes represent the observation time in Barycentric Julian Date (BJD) and relative flux normalized by the average flux.
The red, green and blue bars in panels (a) and (c) indicate
the times of observations by MITSuME ($g$-,$R_C$-, 
and $I_C$-band photometry), HOWPol (low-resolution spectroscopy)
and MALLS (medium-resolution spectroscopy), respectively. 
The magenta and cyan bars in (a) and (c) indicate $g$-band 
photometry with ARCSAT and high-resolution spectroscopy by ARCES at APO.
Up-arrows below the light curve indicate the peak times of flares
detected by our flare detection method.
(b) Light curve of the largest flare on YZ CMi observed with TESS
during the TESS Sector 7 (denoted by a down-arrow in panel a). 
The horizontal axis is the time from the flare peak (T$_0$(BJD)$=2458507.0233$).
(c) Same as (a) but for the light curve around OISTER observations.
(d) Same as (a) but for the light curve around APO observations.
}\label{fig:TESS_LC1}
\end{figure}

\section{Results}

\subsection{TESS observations}
Fig. \ref{fig:TESS_LC1} shows the light curve of YZ CMi observed with TESS.
We can clearly see a sinusoidal modulation with 
a period of $\sim 2.8$ days and many flares.
We applied the Phase Dispersion Minimization(PDM) method \citep{Stellingwerf1978}
to the flare-removed light curve and found that 
the best-estimated period of the sinusoidal modulation 
is $2.774\pm0.0014$ days,
which is consistent with the rotation period of YZ CMi \citep{Pettersen1983,Morin2008}.

\begin{figure}
 \begin{center}
  \includegraphics[width=8cm]{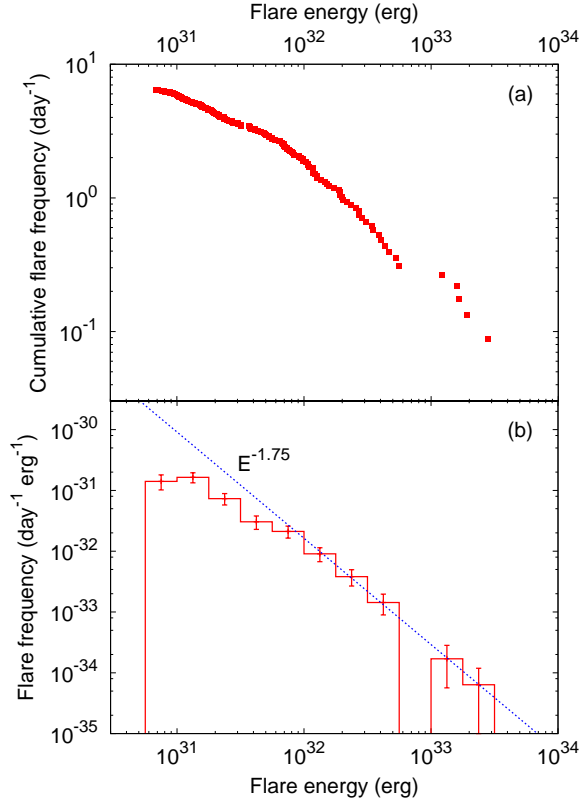} 
 \end{center}
\caption{(a) Cumulative flare frequency distribution as a function of flare energy.
The horizontal axis is the bolometric energy released by flares.
The vertical axis represents the cumulative flare frequency, the number
of flares with the flare energy larger than a give value per day.
Please note that we cannot detect the flares whose occurrence frequency
is much less than $\sim 4.5\times 10^{-2}$ day$^{-1}$ since the total observation time 
is $\sim 22$ days.
(b) Flare frequency distribution as a function of flare energy.
The horizontal axis represents the flare frequency normalized by the bin width.
Dotted line represents a power-law fit to the data
in the flare energy range from $10^{32}$ erg to $10^{34}$ erg.
The power-law index estimated from the fit is $-1.75\pm 0.04$.
}\label{fig:TESS_FFD}
\end{figure}

After removing the rotational light variations, we detected 145 flares
in the TESS light curve. 
As shown in Fig. \ref{fig:TESS_LC1}, the largest flare observed with 
TESS shows a flare peak amplitude of $\sim 14$ \%
relative to the average stellar brightness and a duration 
of $\sim 70$ min.
We estimated the bolometric energy released by flares under 
the assumption that the spectral energy distribution of 
the flare component is similar to
blackbody radiation with an effective temperature of $10000$ K 
(e.g., \cite{Hawley1991,Hawley1992} for flares on M dwarfs).
We also assumed that the effective temperature and radius of YZ CMi are
$3300$ K and $0.3$ R$_{\odot}$ \citep{Gaidos2014,Newton2015,Houdebine2016}.
The amplitude of a flare ($\Delta F/F$) on YZ CMi observed in TESS band ($6000$ -- $10000$ \AA; 
centered on the traditional Cousins $I_C$-band, $\lambda _0 \sim 8000$ \AA; \cite{Ricker2015})
can be written as 
\begin{equation}
\frac{\Delta F}{F} = \frac{A_{\rm flare}}{A_{\rm 1/2\, star}} \frac{\int B(T_{\rm flare}, \lambda) S(\lambda)\, d\lambda}{\int B(T_{\rm star}, \lambda) S(\lambda)\, d\lambda},    
\end{equation}
where $A_{\rm flare}/A_{\rm 1/2\, star}$ is the fraction of flare emitting area normalized by the effective area of stellar hemisphere,
$T_{\rm flare}$ and $T_{\rm star}$ are the effective temperature of flare component
and that of star, $\lambda$ is the wavelength, $B(T,\lambda)$ is the Plank function, 
and $S(\lambda)$ is the spectral response function of TESS detector.
The bolometric luminosity of flare ($L_{\rm flare}$) can be estimated by
\begin{equation}
L_{\rm flare}=\sigma T^4_{\rm flare} A_{\rm flare},
\end{equation}
where $\sigma$ is the Stefan–Boltzmann constant.
For the flare with $\Delta F/F=1$, $A_{\rm flare}/A_{\rm 1/2 star}$ would be $\sim 0.018$
which corresponds to the flare bolometric luminosity of $\sim 2.4\times 10^{31}$ erg s$^{-1}$.
It should be noted that the fraction of bolometric flux for 
the blackbody radiation with an effective temperature of 
$10000$ K which falls within the TESS bandpass 
is only $\sim 20$ \%. 
According to \citet{Kowalski2013}, some flares exhibit large Balmer jump. 
The near-ultraviolet (NUV) continuum for such flares differs from the blackbody 
and the blackbody approximation would underestimate the NUV continuum 
flare flux by a factor of two \citep{Kowalski2019}.
This suggests that the estimated bolometric flare energy may have an
uncertainty by a factor of $2$-$3$.
The bolometric energy released by observed flares ranges from
$7 \times 10^{30}$ erg to $3\times 10^{33}$ erg.
Fig. \ref{fig:TESS_FFD} shows the cumulative flare frequency distribution
(CFFD; panel a) and flare frequency distribution (FFD; $dN/dE$, panel b) 
as a function of flare energy ($E$), respectively.
The FFD can be fitted by a power-law distribution ($dN/dE \propto E^{-\alpha}$) with 
a power-law index of $\alpha = 1.75\pm 0.04$ in the 
bolometric flare energy range of $10^{32}$ to $10^{34}$ erg.
The power-law slope from YZ CMi's FFD is similar to those of 
other M dwarfs (e.g., \cite{Lacy1976,Shakhovskaia1989,Audard2000,Hawley2014,Silverberg2016}) and 
the Sun (e.g., \cite{Crosby1993,Aschwanden2000}).


\subsection{OISTER photometry and spectroscopy}
\begin{figure}
 \begin{center}
  \includegraphics[width=16cm]{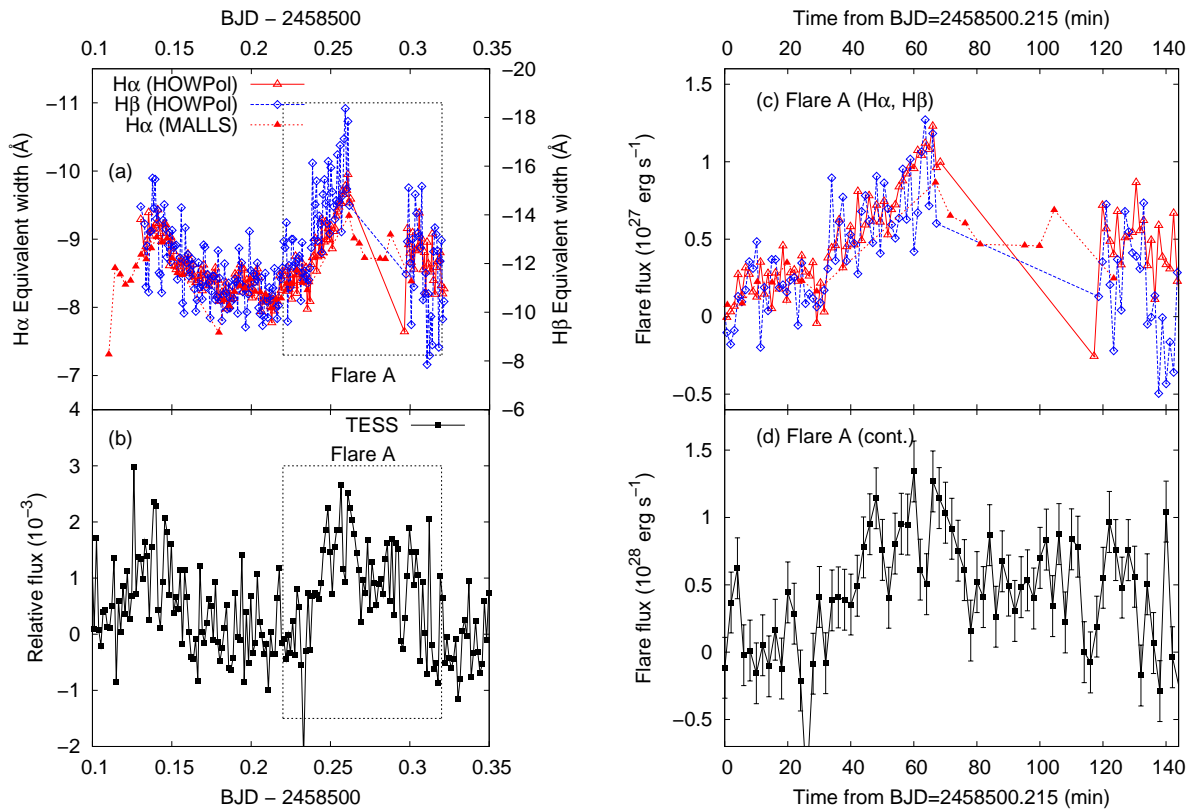} 
 \end{center}
\caption{(a) H$\alpha$, H$\beta$ light curve of YZ CMi on January 16.
The horizontal axis represents
the observation time in Barycentric Julian Date (BJD).
Left and right vertical axes represent
equivalent widths of the H$\alpha$ 
and H$\beta$ lines, respectively (values 
are negative for emission lines).
Open and filled triangles indicate the equivalent width of the H$\alpha$ line
measured from spectra obtained with Kanata/HOWPol and Nayuta/MALLS, respectively.
Open diamonds represent the equivalent width of the H$\beta$ line 
measured from spectra obtained with Kanata/HOWPol.
(b) TESS light curve of YZ CMi during the OISTER observations on January 16.
The vertical axis represents the relative flux normalized by the stellar
average flux.
(c) Enlarged H$\alpha$ and H$\beta$ light curves of flare A.
The horizontal and vertical axis represent
the time from BJD 2458500.215 and the flare component's luminosity.
Open triangles and open diamonds indicate 
the luminosities of the flare component in the H$\alpha$ and H$\beta$ lines.
The flux calibration for the H$\alpha$ and H$\beta$ lines were
performed by using the quiescent spectra taken from \citet{Kowalski2013}, 
$g$- and $R_C$-band magnitudes.
(d) Same as panel (c) but for the flare component's continuum luminosity in the TESS-band ($6000$ -- $10000$ \AA).
}\label{fig:TESS-Halpha-0116}
\end{figure}

\begin{figure}
 \begin{center}
  \includegraphics[width=16cm]{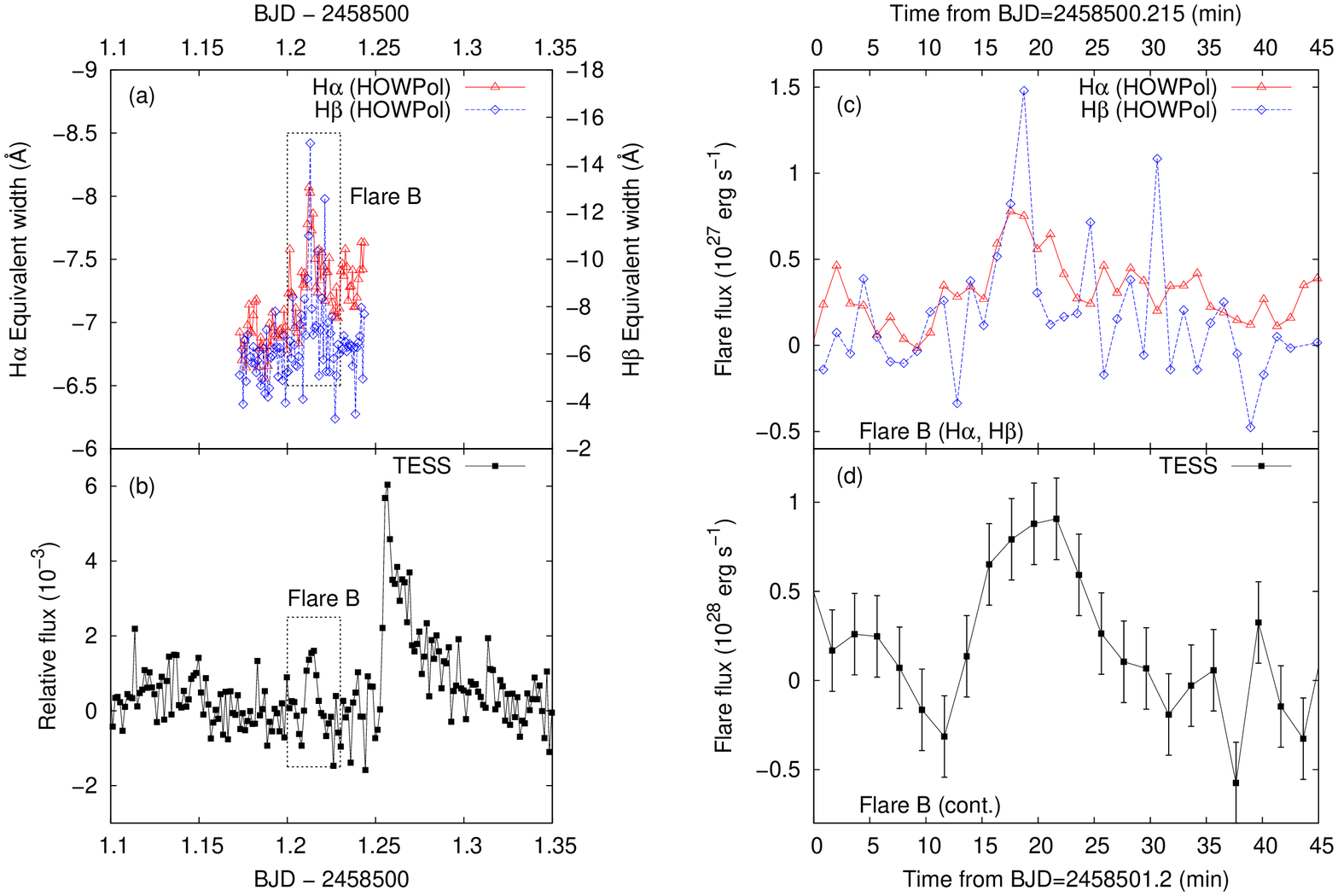} 
 \end{center}
\caption{
The axes and symbols are plotted as done in Fig. \ref{fig:TESS-Halpha-0116}. 
(a) H$\alpha$ light curve of YZ CMi on January 17.
(b) TESS light curve of YZ CMi during the OISTER observations on January 17.
(c) H$\alpha$, H$\beta$ light curve of flare B.
(d) Same as panel (c) but for the flare component's continuum luminosity in the TESS-band.
}\label{fig:TESS-Halpha-0117}
\end{figure}

\begin{figure}
 \begin{center}
  \includegraphics[width=16cm]{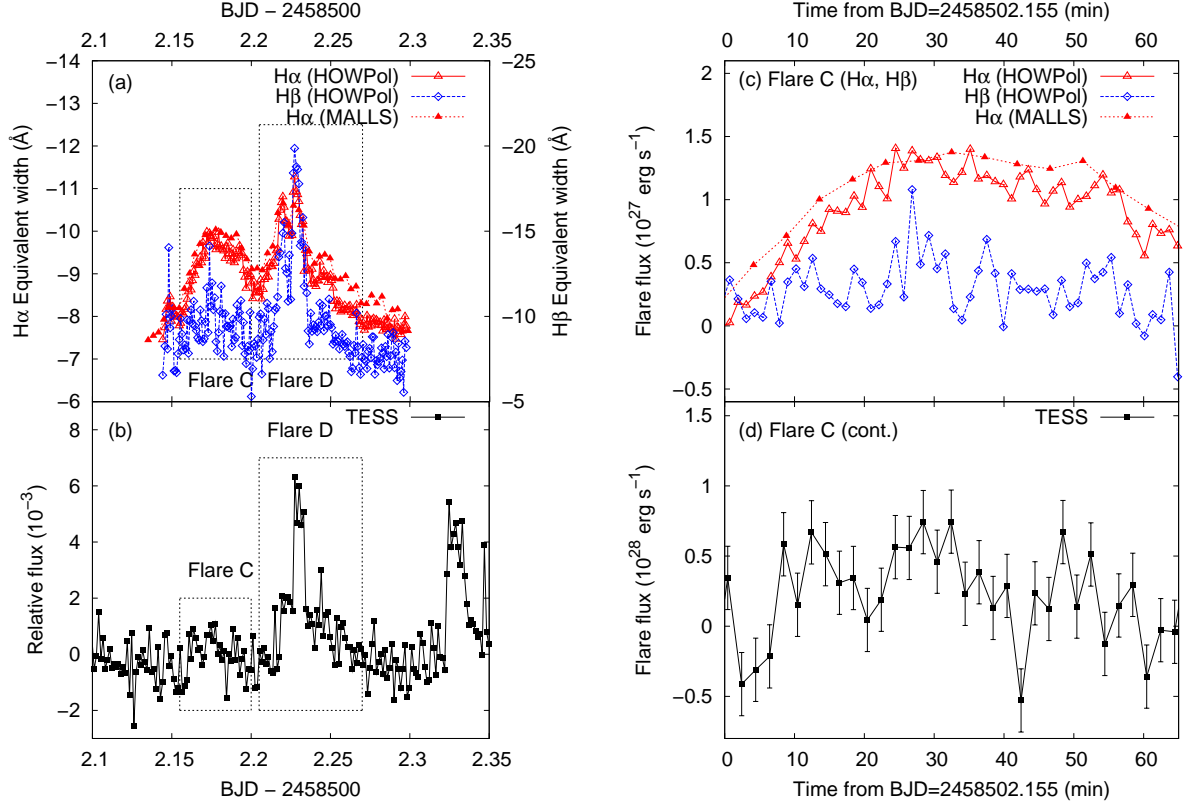} 
 \end{center}
\caption{
The axes and symbols are plotted as done in Fig. \ref{fig:TESS-Halpha-0116}.
(a) H$\alpha$ and H$\beta$ light curves of YZ CMi on January 18.
(b) TESS light curve of YZ CMi during the OISTER observations on January 18.
(c) H$\alpha$ and H$\beta$ light curves of flare C.
(d) Same as panel (c) but for the flare component's continuum luminosity in the TESS band.
}\label{fig:TESS-Halpha-0118}
\end{figure}

 During the OISTER observing campaign, we detected 4 H$\alpha$ flares
from time-resolved spectroscopy as shown in Fig. \ref{fig:TESS-Halpha-0116},
\ref{fig:TESS-Halpha-0117}, and \ref{fig:TESS-Halpha-0118}. 
On January 16, 2019, an H$\alpha$ flare with a peak timing of 2458500.26
(``flare A'')
was detected (Fig. \ref{fig:TESS-Halpha-0116} (a) and (b)).
During this flare, H$\alpha$ and H$\beta$
equivalent widths
\footnote{
The equivalent width of a spectral line is defined as a area
 of the line on a plot of the continuum-normalized intensity
 as a function of wavelength. In this paper, since a negative value 
 of the equivalent width indicates line emission, a decrease
 (negative change) of equivalent width indicates an increase of
 emission line flux.
}
changed by $-2$ \AA \ and $-7$ \AA, respectively.
In addition to the enhancement of the Balmer emission lines, 
the continuum brightness observed with TESS also increased by 0.3\%.
We estimated the luminosity of the flare component by using the distance
 to YZ CMi ($5.99$ pc; \cite{Gaia2018}), 
the flux-calibrated quiescent spectra \citep{Kowalski2013}, 
$g$-, $R_C$-, and $I_C$-band magnitudes.
The peak luminosity of this flare in the H$\alpha$ and H$\beta$ lines
is $1.2\times 10^{27}$ erg  s$^{-1}$ and $0.9\times 10^{27}$ erg s$^{-1}$,
respectively (Fig. \ref{fig:TESS-Halpha-0116} (c)).
The peak luminosity of this flare in the TESS band 
is estimated to be $1.5 \times 10^{28}$ erg s$^{-1}$ 
(Fig. \ref{fig:TESS-Halpha-0116} (d)).


During the short observing run on Janary 17, 2019, 
a small and short-duration H$\alpha$ flare was detected 
at BJD 2458501.2125 (``flare B''). 
The equivalent width of the H$\beta$ emission line and continuum flux
observed with TESS also increased during this flare (Fig.  \ref{fig:TESS-Halpha-0117}).
The amplitude of this flare in H$\alpha$, H$\beta$ and TESS band
are $-1$ \AA\ , $-4$ \AA\ , and 0.2\%, respectively. These values
correspond to the flare peak luminosity of
$0.8\times 10^{27}$ erg s$^{-1}$ ($H\alpha$), 
$\sim 1.2\times 10^{27}$ erg s$^{-1}$  ($H\beta$), and 
$0.9\times 10^{28}$ erg s$^{-1}$ (TESS band), respectively. 

\begin{figure}
 \begin{center}
  \includegraphics[width=16cm]{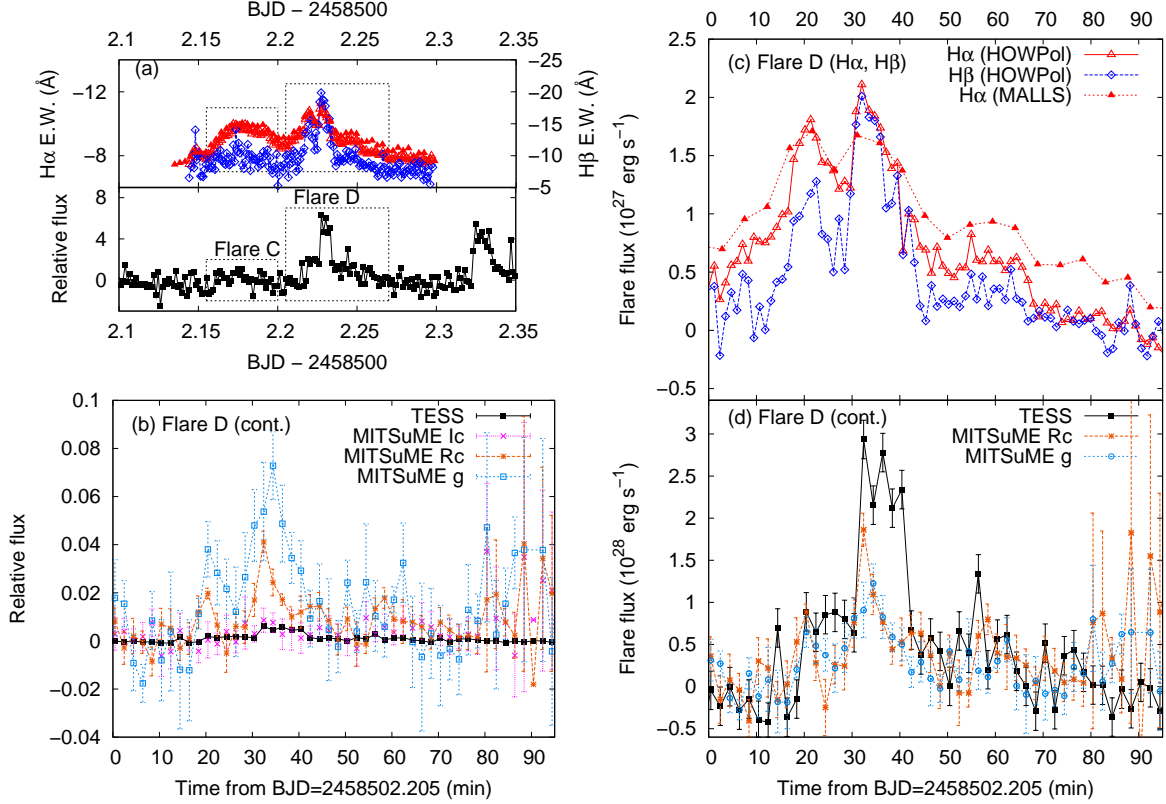} 
 \end{center}
\caption{
(a) H$\alpha$, H$\beta$ and TESS light curves of YZ CMi on January 18.
(b) Light curve of a white-light flare (flare D) on January 18.
Filled squares, crosses, asterisks, and open squares
represent the brightness change in the 
TESS- ($6000$--$10000$ \AA, centered on traditional $I_C$-band; \cite{Ricker2015}) 
$I_C$- ($\lambda _0=7980$ \AA, FWHM$=1540$ \AA; \cite{Bessell2005}), 
$R_C$- ($\lambda _0=6407$ \AA, FWHM$=1580$ \AA; \cite{Bessell2005}), and
$g$-band ($\lambda _0=4770$ \AA, FWHM$=1380$ \AA; \cite{Fukugita1996})
relative to the pre-flare brightness in each band.
Each $I_C$-, $R_C$-, and $g$-band data point and error bar
indicate the average value and standard error derived
from all data points obtained with MITSuME 
during each TESS exposure.
The effective wavelength ($\lambda _0$) and bandwidth
($\Delta \lambda$)
are \ 
for $g$-band \citep{Fukugita1996},
$\lambda _0=6407$ \AA\ and $\Delta \lambda=1580$ \AA\ 
for $R_C$-band \citep{Bessell2005},
$\lambda _0=7980$ \AA\ and $\Delta \lambda=1540$ \AA\ 
for $I_C$-band \citep{Bessell2005}, respectively.
(c) H$\alpha$ and H$\beta$ light curve of flare D.
(d) Same as panel (a) but for the flare component's continuum luminosity in the TESS, $R_C$-, and $g$-band.
}\label{fig:TESS-Halpha-0118-cont.}
\end{figure}

We detected 2 H$\alpha$ flares during the OISTER observation on January 18, 2019.
These two flares show different properties:
the flare at BJD 2458502.177 (``flare C'') shows slow rise and slow decay,
while the double-peaked flare at BJD 2458502.228 (``flare D'') shows rapid rise and
rapid decay.
For flare C, although the amplitude of the H$\alpha$ line
is comparable to those of flare A and B, we cannot find any 
clear brightening in TESS light curve.
The equivalent width of the H$\beta$ line also did not 
exhibit a clear change during this flare.
On the other hand, for flare D, we can see a clear white-light flare
with an amplitude of 0.6\% in TESS band, though the flare amplitude
of the H$\alpha$ line is only $50$\% larger than that of flare C ($3$ \AA).
Moreover, the amplitude of flare D in the H$\beta$ line is $11$ \AA, which
is much larger than that of flare C.
Flare D was also detected by multi-color photometry observed with MITSuME,
as shown in Fig. \ref{fig:TESS-Halpha-0118} (b) and (d).
The amplitudes of flare D in the $g$-, $R_C$-, and $I_C$-bands
are estimated to be 
$6.7$\%, $3.4$\%, and $0.8$\%, respectively, 
by using the MITSuME data around the flare peak 
in TESS band
(from BJD 2458502.227 to 2458502.229).
The blackbody fitting to the SED of the flare component around the peak
derived from the $g$-, $R_C$-, and TESS band data yields an effective temperature of 
$5900 \pm 1000$ K, 
which is comparable to the effective temperature of solar
white-light flares (e.g., \cite{Watanabe2013},\cite{Kerr2014}, \cite{Kleint2016}) and 
$3000$ - $4000$ K lower than the typical effective temperature of white-light flares 
on M dwarfs (e.g., \cite{Hawley1991,Hawley1992,Hawley2003,Kowalski2013}).
According to \citet{Kowalski2019}, the effective temperature of
flare component estimated from the optical continuum with a wavelength $\lambda > 4000$ \AA\ 
tends to be low ($\sim 6000$ K) for flares exhibiting large Balmer jump ratio.
\citet{Kowalski2013} reported that flares with smaller peak amplitude 
and longer flare FWHM (full width at half maximum) time
tend to show a larger Balmer jump ratio than impulsive flares.
As shown in Fig. \ref{fig:TESS-Halpha-0118-cont.}, flare D is not a impulsive flare
because of the small peak amplitude ($7.6$ \% in $g'$-band) and long flare FWHM time ($\sim 10$ min).
This morphological property of light curve for flare D is 
similar to that for flares showing a low-temperature
continuum at $\lambda > 4000$ \AA.

For the flare C, the peak flux of flare component
in the H$\alpha$ line is $1.4\times 10^{27}$ erg s$^{-1}$ (Fig.  \ref{fig:TESS-Halpha-0118} (c)), which
is comparable to that of flare D ($2.1\times 10^{27}$ erg s$^{-1}$;
Fig. \ref{fig:TESS-Halpha-0118-cont.} (c)).
However, the luminosity of flare component in the TESS-band continuum
at the peak timing of flare C is $<6\times 10^{27}$ erg s$^{-1}$
 (Fig. \ref{fig:TESS-Halpha-0118} (d)).
This value is $<1/5$ of the flare D's peak luminosity
in the TESS-band (Fig. \ref{fig:TESS-Halpha-0118-cont.} (d)).

%

The flare energies released in the optical continuum (TESS band), H$\alpha$, 
and H$\beta$ emission lines for each flare detected by OISTER observations
are summarised in Table \ref{table:OISTER_flares}.
Among these flares, only flare C did not show a white-light flare.
For other flares associated with white-light flares, 
the flare energy released in the 
TESS-band continuum ranges from 
$3\times 10^{30}$ erg to $3.6\times 10^{31}$ erg.
The ratio of the flare energy released in the TESS-band continuum to that released
in H$\alpha$ and H$\beta$ lines ranges from $\sim 5$ to $\sim 20$.

\begin{table}
  \tbl{List of flares detected by OISTER observations}
  {
\begin{tabular}{ccccccccccc}\hline
 & & \multicolumn{4}{c}{Peak luminosity ($10^{27}$ erg s$^{-1}$)}  & \multicolumn{4}{c}{Energy ($10^{31}$ erg)} & Duration (min) \\
Flare ID & peak (BJD)\footnotemark[$*1$] &  $L_{\rm TESS}$\footnotemark[$*2$] & $L_{\rm bol}$\footnotemark[$*3$] & $L_{{\rm H}\alpha}$\footnotemark[$*4$] & $L_{{\rm H}\beta}$\footnotemark[$*4$]  & $E_{\rm TESS}$\footnotemark[$*2$] & $E_{\rm bol}$\footnotemark[$*3$] & $E_{{\rm H}\alpha}$\footnotemark[$*4$] & $E_{{\rm H}\beta}$\footnotemark[$*4$] & $\tau _{\rm flare}$ \footnotemark[$*1$] \\
 \hline
A & 2458500.260 & $14$ & $73$ & $1.2$ & $0.8$ & $2.9$ & $15$ & $0.25$ & $0.3$ & $50$\\ 
B & 2458501.212 & $9$ & $48$ & $0.8$ & $\sim 1.2$ & $0.3$ & $1.8$ & $0.06$ & $0.05$ & $15$ \\
C & 2458502.177 & $<6$ & $<32$ & $1.4$ & $<0.5$ & $<0.4$ & $<2.1$ & $0.47$ & $<0.2$ & $>70$\footnotemark[$*5$]  \\
D & 2458502.228 & $29$ & $155$ & $2.1$ & $2.1$ & $3.6$ & $18$ & $0.18$ & $0.24$ & $35$ \\
         \hline
  \end{tabular}
  }\label{table:OISTER_flares}
      \begin{tabnote}
 \footnotemark[$*1$] The flare duration and peak time were measured from the H$\alpha$ light curve obtained with Kanata/HOWPol.
  \footnotemark[$*2$] Luminosity and energy emitted in the TESS bandpass ($6000$ -- $10000$ \AA).
  \footnotemark[$*3$] We assumed that 
      the effective temperature of the flare component, temperature and radius of YZ CMi 
      are $10^4$ K, $3300$ K, and $0.3 {\rm R}_{\odot}$.
 \footnotemark[$*4$] The flux calibration for the H$\alpha$ and H$\beta$ lines were performed
 by using the quiescent spectra taken from \citet{Kowalski2013}, $g$- and $R_C$-band magnitudes.
 \footnotemark[$*5$] The flare D started before the flare C ended.
 \end{tabnote}
\end{table}


\section{Discussion}
\subsection{White-light and non-white-light flares}
In this subsection, we focus on two H$\alpha$ flares (C and D)
observed on January 18, 2019.
As shown in Fig. \ref{fig:TESS-Halpha-0118}, the flare C showed 
a clear increase in the H$\alpha$ emission line, but
no significant flare was observed in the continuum 
(Fig. \ref{fig:TESS-Halpha-0118}).
In contrast, flare D showed clear brightening not only
in Bamler lines (H$\alpha$ and H$\beta$) 
but also in the continuum ($g$, $Rc$, $Ic$, and TESS bands).
H$\alpha$ and H$\beta$ light curves for flare D show
two peaks corresponding to two peaks of white-light flare
as shown in Fig. \ref{fig:TESS-Halpha-0118-cont.}.
The H$\alpha$ emission line flux at the peak time of flare C
is $1.4\times 10^{27}$ erg s$^{-1}$, which is comparable to that
at the peak of flare D ($2.1\times 10^{27}$ erg s$^{-1}$).
Although the H$\alpha$ line fluxes at flare peaks of C and D
are roughly the same,
the continuum flux in TESS band at the peak time of flare D
($2.9 \times 10^{28}$ erg s$^{-1}$) is one order of magnitude 
greater than that of flare C ($<6\times 10^{27}$ erg s$^{-1}$).

The differences between flare C and D are
the time scale of flare and 
the ratio of the H$\alpha$ line flux to the continuum flux at the flare peak.
The rise time of flare C is much longer than that of that of flare D.
In the case flare C, the equivalent width of H$\alpha$ emission
gradually increased over $\sim 25$ minutes and decayed over $>40$ minutes.
The ratio of the H$\alpha$ line flux to the continuum flux in TESS band
was $>0.2$
for the flare C.
In contrast, the rise time of flare D is $\sim 5$ minutes,
which is comparable to that of flare D observed in continuum light.
The peak timings of flare D in H$\alpha$, H$\beta$, and continuum
are the same within the time-resolution of observations.
The ratio of the H$\alpha$ line flux to the continuum flux in TESS band
was $\sim 0.07$ for the flare D.
According to \cite{Kowalski2013}, the ratio of the H$\alpha$
emission flux to the total flux at the flare peak ranges from
$0.0005$ to $0.10$ for various types of flares on M dwarfs.
The ratio of the H$\alpha$ line flux to the total flux 
at the peak time of the gradual flares showing slow 
rise/decay ($L_{{\rm H}\alpha}/L_{\rm bol.} \sim 3$-$10$ \%) 
tends to be larger than that of impulsive flares ($<1.7$ \%).
The difference in the flux ratio of H$\alpha$ and continuum
between the flare C and flare D agree with this tendency.

According to 1-dimensional radiative hydrodynamic model 
calculation of stellar flares on M dwarfs by \citet{Namekata2020} 
using the RADYN code \citep{Carlsson1997,Allred2015},
the H$\alpha$ line and continuum intensity increase 
as the energy deposition rate increases and the correlation between 
the intensity of H$\alpha$ line ($I_{\rm H \alpha}$) 
and that of continuum ($I_{\rm cont.}$) can be expressed by 
$I_{\rm cont.} \propto I_{\rm H \alpha} ^{0.5}$. 
This suggests that the fraction of continuum intensity 
relative to the H$\alpha$ intensity increases as the 
continuum intensity increases. In the case of the flare
caused by the lager non-thermal electron flux,
the ratio of the continuum intensity to the H$\alpha$ intensity is much larger
than that for the flare caused by the smaller non-thermal electron flux. 
This non-linear correlation between the intensity of H$\alpha$ line and that of 
continuum suggests that the observed difference in the ratio of the continuum 
flux to the H$\alpha$ flux between the flare C and flare D can be 
explained by the difference in 
the energy deposition rates. 
This interpretation is also consistent with the fact that the 
rise time
of flare C (small deposition rate) is longer than that of flare D (large deposition rate).
The long rise time of flare C suggest that flare C
is actually a group of many small flares
for which detectable white light emission is not expected.
In addition to the continuum flux, the H$\beta$ line flux at the peak
time of flare C is also smaller than that at the peak time of flare D.
The ratio of H$\beta$ line flux to the H$\alpha$ line flux (H$\beta$/H$\alpha$)
is $<0.4$ for flare C and $\sim 1.0$ for flare D.
H$\beta$/H$\alpha$ depends on the electron density, temperature and optical depth
according to \citet{Drake1980}.
Since the optical depth of the H$\beta$ line is smaller than that of the H$\alpha$,
the H$\beta$ line flux may behave like the continuum
rather than the H$\alpha$ line flux during the flare.
The smaller value of H$\beta$/H$\alpha$ for flare C than that for flare D
suggests that the electron density of the flare region for flare C
is smaller than that for flare D.
This may be consistent with the non-thermal electron flux for flare 
C being smaller than that for flare D.

As mentioned above, 
although the intensity of H$\alpha$ line from the flare region for flare C 
is expected to be smaller than that for flare D, 
the peak H$\alpha$ luminosity of flare C is comparable to that of flare D.
This may suggest that the area of flare region
for flare C is larger than that for flare D.
As we discuss in later section, the larger area of flare region
for flare C would lead to the longer flare duration.
Solar non-white light flares have longer soft X-ray flare durations than
white-light flares with the similar GOES X-ray class
(e.g., \cite{Watanabe2017}). 
This tendency is consistent with the differences in the duration and peak continuum flux
between the flare C and flare D.

\subsection{H$\alpha$ line profile changes during flares C and D}
\begin{figure}
 \begin{center}
  \includegraphics[width=14cm]{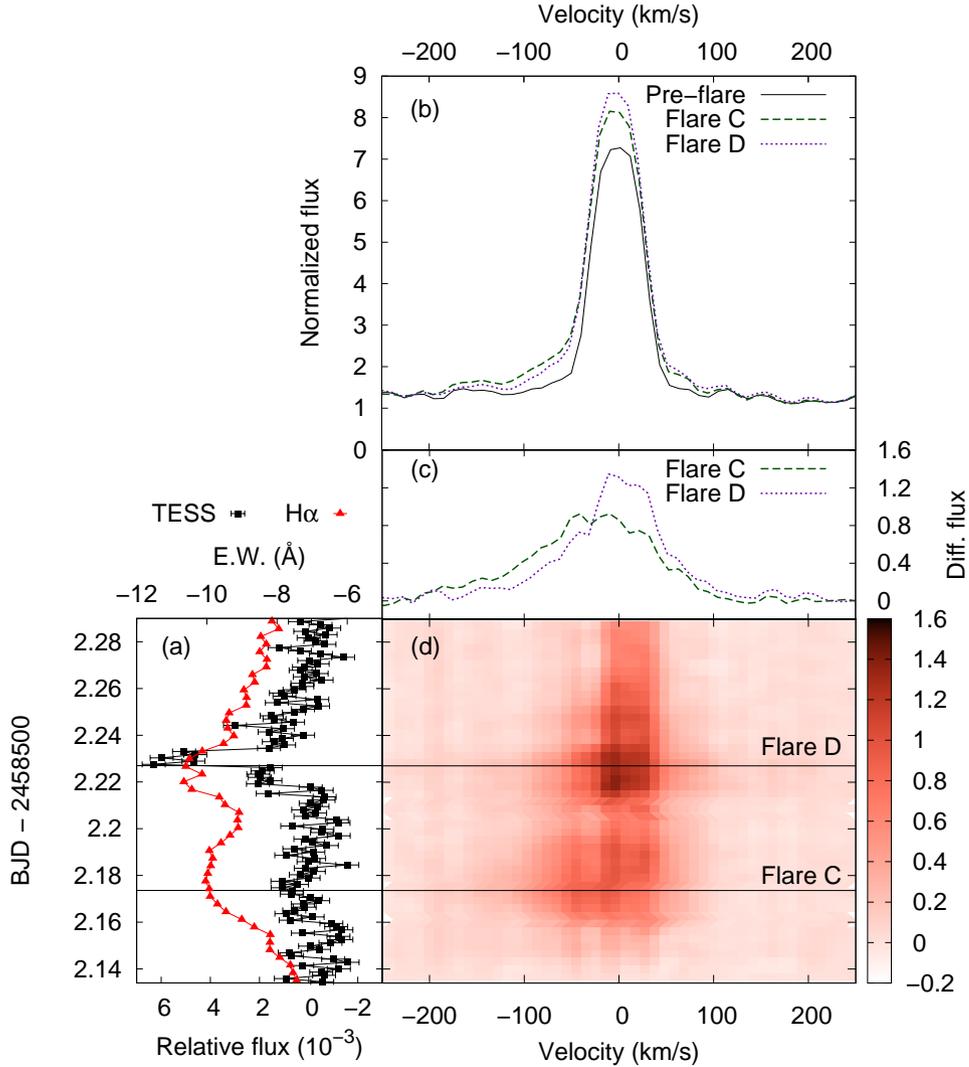} 
 \end{center}
\caption{
(a) Light curve of YZ CMi during flare C and D.
Filled squares (black) and triangles (red) represent 
continuum flux relative to the average flux observed
with TESS and equivalent width of H$\alpha$ line observed
with Nayuta/MALLS.
The vertical axis indicates the time of observation
in Barycentric Julian Date (BJD).
The lower and upper horizontal axes represent
continuum flux relative to the average stellar flux
and equivalent width in unit of \AA\ , respectively.
(b) Line profile of H$\alpha$ emission line.
The horizontal and vertical axes represent
the Doppler velocity from H$\alpha$ line center and 
flux normalized by the continuum. Solid (black), dashed (green), and,
dotted (purple) lines indicate the line profile
before flares (BJD 2458502.135), that at the peak timing of
flare C (BJD 2458502.177) and that of flare D (BJD 2458502.227).
(c) Same as (b), but for the line profile changes from the pre-flare
to the peak times of flare C and D.
(d) Time evolution of H$\alpha$ line profile.
The horizontal and vertical axes represent the 
Doppler velocity from H$\alpha$ line center and
time of observation. The color map represents
the line profile changes from the pre-flare profile. 
}\label{fig:Ha-3dmap}
\end{figure}

\begin{figure}
 \begin{center}
  \includegraphics[width=8cm]{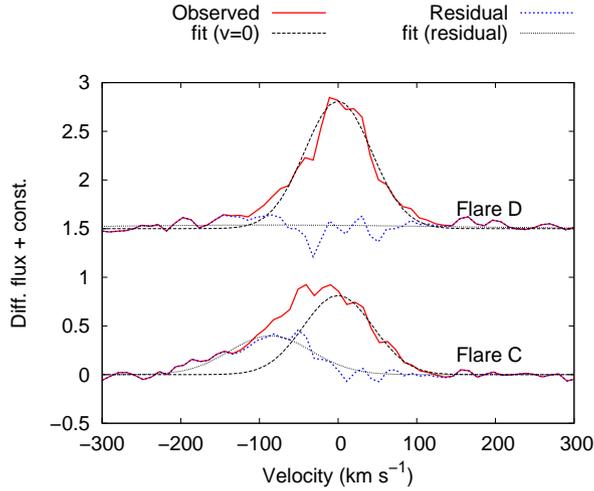} 
 \end{center}
\caption{
 H$\alpha$ line profile changes from the 
 pre-flare (BJD 2458502.135) to the peak times of 
 flare C (BJD 2458502.177; lower) 
 and D (BJD 2458502.227; upper).
 Red sold lines indicate the observed line profile changes.
 Black dashed lines represent a Gaussian fit assuming the line-of-sight
 velocity of 0 km s$^{-1}$ to the red-part ($>0$ km s$^{-1}$) of
 the observed line profile changes.
 Blue dotted lines indicate residuals between the observed line profile changes and the Gaussian fits (thin-dashed lines).
 Black dash-dotted lines represent a Gaussian fit to the
 residuals (blue dotted lines).
 The line-of-sight velocity of the residual component (blue enhancement) 
 at the peak time of the flare C is $-85\pm 3$ km s$^{-1}$.
}\label{fig:Flare_CD_fit}
\end{figure}

\begin{figure}
 \begin{center}
  \includegraphics[width=8cm]{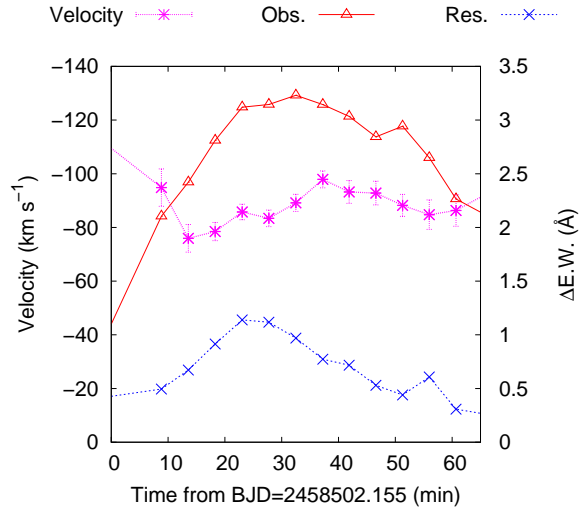} 
 \end{center}
\caption{
 Time variation of the line-of-sight velocity and equivalent
 width of the blue-shifted enhancement during the flare C.
 Magenta asterisks and blue crosses indicate the velocity
 and equivalent width of blue-shifted excess component.
 Red open-triangles indicate the equivalent width of 
 flare component (change in the equivalent width from the pre-flare state).
}\label{fig:Flare_C_velocity}
\end{figure}

During flare C, the profile of H$\alpha$ line showed blue asymmetry
(Fig. \ref{fig:Ha-3dmap}).
As shown in Fig. \ref{fig:Flare_CD_fit}, the velocity of the blue-shifted
excess component at the peak time of flare C is $-85\pm 3$ km s$^{-1}$.
The blue-shifted excess component with the line of sight velocity of 
$-80$ -- $-100$ km s$^{-1}$ has been seen for 60 minutes during flare C (Fig. \ref{fig:Flare_C_velocity}).
Around the peak time of flare C, roughly $1/4$ - $1/3$ of the H$\alpha$ emission
of flare component was emitted from the blue-shifted excess component.
On the other hand, no clear asymmetry in the H$\alpha$ line profile was observed
during the flare D (Fig. \ref{fig:Ha-3dmap} and \ref{fig:Flare_CD_fit}).

Since the velocity of blue-shifted component observed during flare C 
is one order of magnitude larger than the stellar rotation 
velocity ($v\sin i \sim 6$ km s$^{-1}$; e.g., \cite{Houdebine2016}),
observed blue asymmetry cannot be explained by the rotationally 
modulated emission from the co-rotating prominence (e.g., \cite{CollierCameron1989}).

According to \citet{Fisher1985}, when the non-thermal electron flux
is low, a weak chromospheric evaporation with the upward velocity 
$<30$ km s$^{-1}$ occurs (gentle evaporation).
The threshold for the gentle evaporation depends on the
energy spectra of non-thermal electron beam (low-energy cutoff and power-law index; \cite{Fisher1989}).
As discussed above, the weak non-thermal electron flux is suggested from 
the weak continuum intensity relative to the H$\alpha$
intensity for the flare C. 
The observed velocity of blue asymmetry during flare C ($80$-$100$ km s$^{-1}$)
is a few times larger than the theoretical prediction for solar flares
and that of blue asymmetry 
observed during the gradual phase of solar flares (e.g., \cite{Schmieder1987}).

The Doppler velocity of observed blue asymmetry is
comparable to that of blue asymmetry observed in chromospheric lines
during the initial phase of solar flares, which is proposed  to be 
caused by the cool plasma lifted up by the expanding hot plasma
(e.g. \cite{Tei2018}, \cite{Li2019}). However, the duration of these blue
asymmetries (a few min) are roughly two orders  
of magnitude shorter than
that of the blue asymmetry observed during flare C ($\sim 60$ min).
\citet{Honda2018} reported the similar long-duration H$\alpha$ flare
showing blue asymmetry in the H$\alpha$ line profile on the M4.5 dwarf EV Lac.
The Doppler velocity of blue-shifted excess component for this event
is $\sim 100$ km s$^{-1}$ and the blue asymmetry has lasted for $>2$ hours.
Since there were no high-precision photometric data for 
the H$\alpha$ flare showing a blue asymmetry reported by \citet{Honda2018}, 
it is unclear whether the typical long duration H$\alpha$ flares
with blue asymmetry are non-white-light flares as flare C or not.

The observed velocity of the blue asymmetry during flare C
is also comparable to the velocity of H$\alpha$ surges 
(e.g. \cite{Canfield1996}) 
and that of prominence/filament eruptions
(e.g., \cite{Gopalswamy2003}) of the Sun.
In case of stellar flares, since we cannot obtain spatial information
of stellar surface, such eruptions and surges may also be possible causes of
the blue asymmetry in the H$\alpha$ line associated with flares.
\citet{Vida2016} reported several flares on the M4 dwarf V374 Peg exhibiting 
blue asymmetry whose Doppler velocity ranges $200$ to $400$ km s$^{-1}$.
The duration of these blue asymmetries ranges from $10$ to $30$ minutes.
The similar high-velocity blue asymmetry has been observed during a flare
on AT Mic \citep{Gunn1994}.
Such high-velocity blue asymmetries during stellar flares are considered to 
cause stellar CMEs (e.g., \cite{Vida2016}).
According to \citet{Gopalswamy2003}, the average velocity of the core of CMEs
($\sim 350$ km s$^{-1}$)
and average CME velocity ($\sim 610$ km s$^{-1}$)
are $\sim 4$ and $\sim 8$ times larger than that of the associated
prominences ($\sim 80$ km s$^{-1}$), respectively.
This indicates that
the most prominences accelerate as they erupt.
If the similar acceleration mechanisms would work on YZ CMi, 
although the velocity of prominence eruption ($\sim 100$ km  s$^{-1}$)
is smaller than the escape velocity at the stellar surface
($\sim 600$ km s$^{-1}$ for YZ CMi),
the upward-moving material would be accelerated to $\sim 400$-$800$  km  s$^{-1}$.
Since this value is larger than the escape velocity at $\sim 2$-$3$ R$_{\rm star}$
($\sim 350$-$420$ km s$^{-1}$), even the prominence eruption with the velocity
of $\sim 100$ km s$^{-1}$ could cause a CME.
However, it is still unclear whether the prominence eruptions on M
dwarfs can cause the stellar CMEs. 
\citet{Crosley2018} performed optical and low-frequency radio
observations of the M4 dwarf EQ Peg and found no signature of 
type II bursts, which are believed to be excited by shocks driven by CMEs,
during the observed flares.
One interpretation for the absence of type II bursts
is the magnetic suppression of CMEs in active stars.
Numerical studies by \citet{Drake2016} and \citet{Alvarado-Gomez2018}
suggest that if the flare region has strong overlying magnetic fields,
CMEs will be suppressed.
According to \citet{Morin2008}, rapidly-rotating mid-M ($\sim$ M4) dwarfs 
such as YZ CMi have mainly axisymmetric large-scale poloidal fields.
In the case of YZ CMi, the magnetic energy 
in dipole mode accounts for $\sim 70$ percent of the whole magnetic energy, and 
such large-scale and strong dipole magnetic fields 
may cause the suppression of CMEs.
Another interpretation for the absence of stellar type II radio burst
is that CMEs propagating in the corona/wind of active M dwarfs
are “radio-quiet” for ground-based instruments.
\citet{Mullan2019} proposed that due to the large Alfv$\acute{\rm e}$n speed 
in the corona of active M dwarfs, CMEs from these stars could
not satisfy the conditions for the generation of type II radio bursts.
Accroding to 3D magnetohydrodynamic simulations by\citet{Alvarado-Gomez2020},
while CMEs from active M dwarfs can generate shocks in the corona, 
type II radio burst frequencies are down to the ionospheric cutoff and 
therefore it is difficult to detect type II radio bursts from active
M dwarfs by ground-based instruments.

As shown in Fig. \ref{fig:Flare_C_velocity}, the velocity of the blue-shifted
excess component is almost constant ($80$ -- $100$ km s$^{-1}$) during the flare C
and the H$\alpha$ line flux from the blue-shifted component decreases as the later
part of the flare C. 
In the case of solar CMEs, a prominence forms the core of a CME
(e.g., \cite{Gopalswamy2015}) and the prominence material can be 
observed in the H$\alpha$ line
only at early stages.
If the prominence on YZ CMi forms the core of a CME
and H$\alpha$ emission from the prominence would disappear
as the height of the prominence increases, 
then the observed blue-shifted
excess would come from the prominence at the lower layer
and the velocity of the blue asymmetry
would not change so much during the flare C. 
The intensity of the 
blue-shifted excess may decrease as the mass of the cool material
at lower layer decreases.
Another interpretation for the persistent blue-asymmetry with a constant
velocity for $\sim 1$ hour associated with flare C is  that it is 
caused by the ongoing magnetic reconnection and plasma evaporation 
in the multi-threaded coronal loops (e.g. \cite{Warren2006}).
If upward velocities of evaporated plasma in each coronal loop would be
roughly the same, the velocity of blue-shifted excess would also be
constant during the flare.
However, since our observations are limited (only optical H$\alpha$ line
is used here), 
it is difficult to distinguish whether the observed blue asymmetry was  
caused by a prominence eruption, which would cause a stellar CME,
or by the chromospheric evaporation in coronal loops.
Time-resolved and high-resolution X-ray spectroscopy of stellar flares
(e.g., \cite{Argiroffi2019}), 
EUV observations for the coronal dimming associated with stellar flares
(e.g., \cite{Jin2020}),
and radio observations for stellar type II bursts (e.g., \cite{Crosley2018})
simultaneously with optical spectroscopy 
are necessary to investigate 
the connection between the blue asymmetries in the chromospheric 
lines and stellar CMEs. 

\begin{figure}
 \begin{center}
  \includegraphics[width=8cm]{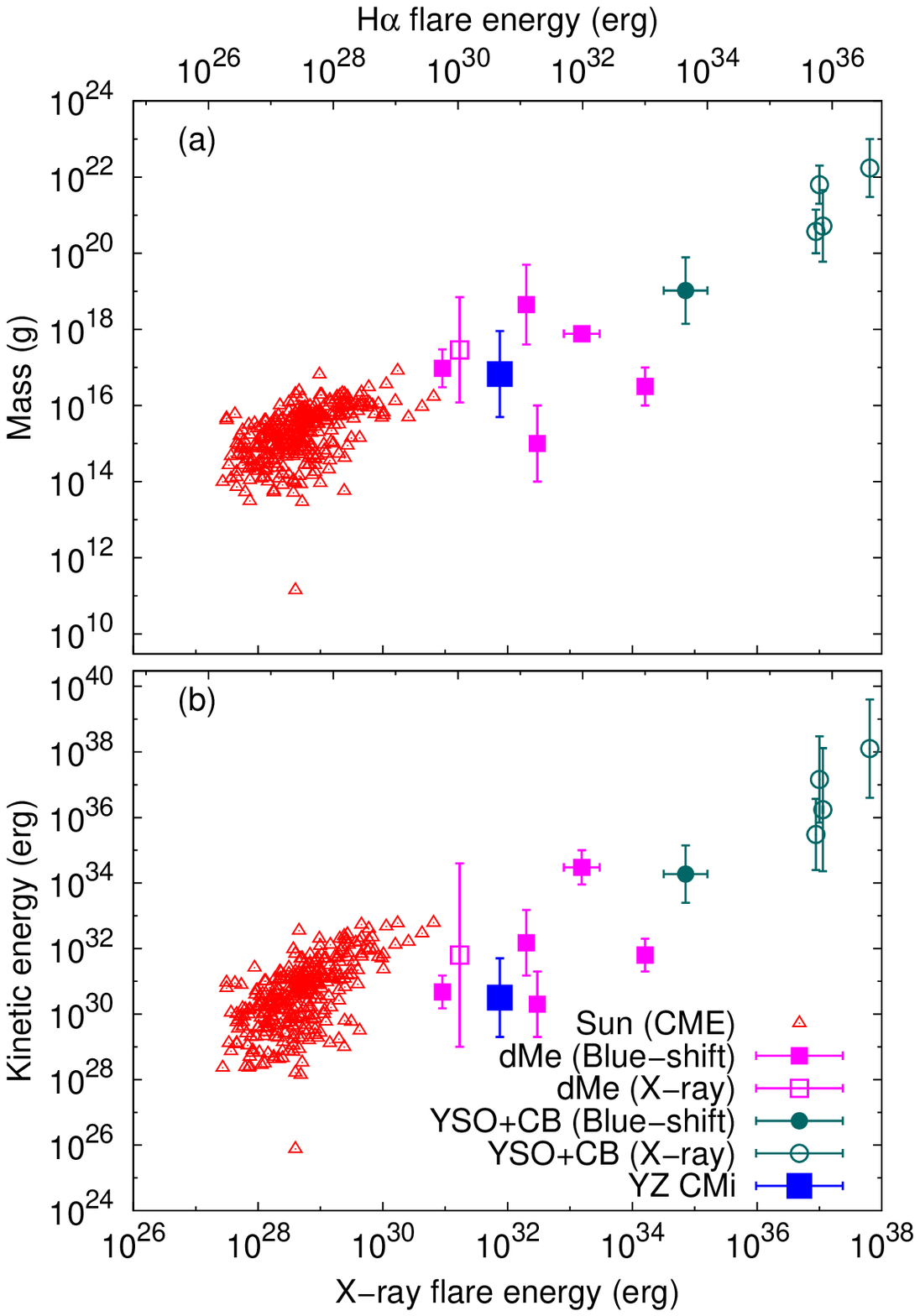} 
 \end{center}
\caption{
(a) Mass of CMEs/prominences as a function of flare energy.
Upper and lower horizontal axis represent the flare energies
emitted in the H$\alpha$ emission line and in X-ray 
(GOES $1$--$8$ \AA\ band), respectively.
The H$\alpha$ flare energy is converted to 
the GOES X-ray flare energy by using equation (\ref{eq:LX2LHa}).
Open triangles (red) represent the CME mass of solar events taken from
\citet{Yashiro2009}. X-ray energy of associated solar flares are derived
from the X-ray fluence in GOES $1$--$8$ \AA band.
Filled squares (magenta) and filled circles (cyan green)
indicate the CME/prominence 
mass of stellar events on M-dwarfs (dMe), young stellar 
objects (YSO) and close binary systems (CB) estimated from
the blue-shifted enhancement of chromospheric lines.
Open squares and filled circles also indicate
the mass of stellar events on M-dwarfs, young stellar 
objects and close binary systems estimated from
the X-ray dimming. Mass and X-ray energy for these events
are taken from \citet{Moschou2019}.
Large filled square (blue) indicates the mass of 
upward-moving material for the flare C on YZ CMi.
The upper and lower error bars for the flare C on YZ CMi represent
the mass of upward-moving material estimated in 
(\ref{cmemass_upper}) and (\ref{cmemass_lower}). 
(b) Same as (a) but for the Kinetic energy of 
CMEs/prominences as a function of flare energy.
}\label{fig:CMEmass}
\end{figure}

As shown in Fig. \ref{fig:Flare_C_velocity}, the equivalent width
of the H$\alpha$ emission from the blue-shifted excess
is $\sim 1$ \AA, 
which corresponds to the luminosity of H$\alpha$ emission of
$L_{{\rm H}\alpha} \sim 6\times 10^{26}$ erg s$^{-1}$.
According to the Non-LTE model of the solar prominence 
by \citet{Heinzel1994}, the flux of the H$\alpha$ line ($F_{{\rm H}\alpha}$)
is roughly $10^{5.5}$ erg s$^{-1}$ cm$^{-2}$ sr$^{-1}$,
if we assume that the optical thickness of the H$\alpha$ line 
($\tau _{{\rm H}\alpha}$) ranges from $1$ to $10$.
If we assume the physical parameters of the prominence on YZ CMi
are similar to that on the Sun, the observed luminosity of H$\alpha$ emission
suggests that the emitting area of blue-shifted H$\alpha$
emission ($A$) would be $\sim 3\times 10^{20}$ cm$^{2}$, which is roughly 10\%
of the area of visible stellar surface.
This value is comparable to the area of starspots estimated from
the amplitude of rotational variations.
By adopting the electron density ($n_e$) - H$\alpha$ line flux ($F_{{\rm H}\alpha}$) correlation and the correlation between 
the electron density ($n_e$) and total Hydrogen density ($n_H$) 
from \citet{Heinzel1994}, 
$n_e \sim 10^{10.5}$ -- $10^{11}$ cm$^{-3}$ for $F_{{\rm H}\alpha} \sim 10^{5.5}$ erg s$^{-1}$ cm$^{-2}$ sr$^{-1}$
and $n_{H} \sim 10^{11}$ cm$^{-3}$.
Although the H$\alpha$ emission is likely from a region whose width along the line of sight is 
consistent with an optical depth of unity 
(which would be much smaller than the prominence size), 
in order to estimate the upper limit of prominence mass, 
we assumed that the geometrical thickness of the upward-moving prominence is comparable to the size of the prominence ($A^{0.5}$).
Under this assumption, the total mass of the upward-moving material 
($M$) can be written as the following form:
\begin{equation}
M\sim A^{1.5} n_H m_H \sim (3\times 10^{20})^{1.5}\ 10^{11}\ 1.7\times 10^{-24} \sim 10^{18}\ {\rm (g)},
\label{cmemass_upper}
\end{equation}
where $m_H$ is the mass of Hydrogen atom.
In the case of solar prominence, the Hydrogen column density ($N_H$) is
estimated to be approximately $1$ -- $5\times 10^{19}$ cm$^{-2}$
(e.g., \cite{Gilbert2005} for eruptive prominence; \cite{Landi2013}
for prominence blobs; \cite{Heinzel2008} for quiescent prominence).
According to \citet{Dunstone2006}, the column density of 
the co-rotating prominence on the K3 dwarf BO Mic (= HD 197890)
is estimated to be $N_H \sim 3\times 10^{19}$ cm$^{-2}$.
If we assume the column density of the upward-moving prominence is roughly
the same order of magnitude with those of the solar and stellar
promineces, the mass of the prominence can be estimated to
be
\begin{equation}
M \sim A N_H m_H \sim 3\times 10^{20}\ (1{\rm -}5) \times 10^{19} \ 1.7\times 10^{-24} \sim  (0.5{\rm -}2.5) \times 10^{16}\ {\rm (g)}.
\label{cmemass_lower}
\end{equation}
The estimated mass of the upward-moving material is the same order 
of magnitude with those estimated by \citet{Vida2016} (V374 Peg)
and \citet{Moschou2019} (DENIS 1048-39 and AD Leo).
Since the observed upward velocity is $80$ -- $100$ km s$^{-1}$,
the kinetic energy of the upward-moving material is estimated to
be the order of $10^{29.5}$ -- $10^{31.5}$ erg, which is roughly
the same order of magnitude with the energy emitted in 
TESS-band continuum ($<4\times 10^{30}$ erg) 
or in the H$\alpha$ line ($5\times 10^{30}$ erg).
According to \citet{Moschou2019}, the luminosity in GOES X-ray band ($1$--$8$ \AA\ ; $L_{X, GOES}$)
can be estimated from the H$\alpha$ luminosity ($L_{{\rm H}\alpha}$) by using the following
linear relation:
\begin{equation}
    L_{X, GOES} = 16 L_{{\rm H}\alpha}. \label{eq:LX2LHa}
\end{equation}
Then the X-ray flare energy of the flare C in the 
GOES $1$--$8$ \AA~
band is estimated to be $8\times 10^{31}$ erg
if we assume the X-ray light curve shape is similar to
that of H$\alpha$.
In the case of X-class solar flares, the X-ray energy in the GOES $1$--$8$ \AA~band
is roughly $1/20$ of the total energy radiated from the
soft X-ray (SXR) emitting plasma \citep{Emslie2012}.
According to \citet{Emslie2012} and \citet{Aschwanden2017}, 
the kinetic energy of solar CMEs is $1$ -- $10$ times greater than
the total energy radiated from the SXR-emitting plasma.
In the case of flare C, the kinetic energy ($10^{29.5}$ -- $10^{31}$ erg) 
estimated from 
the blue-asymmetry in H$\alpha$ line is at least $\sim 2$ order of magnitude
smaller than the thermal energy ($\sim 10^{33}$ erg) estimated from 
the energy emitted in H$\alpha$ line. 
As listed in Table \ref{table:OISTER_flares},
the upper limit of flare energy emitted in the TESS-band continuum
for flare C is comparable to or smaller than that
emitted in the H$\alpha$ line and two orders of magnitude
smaller than the total SXR energy. These suggest that 
the upper limit of bolometric energy for flare C 
listed in Table \ref{table:OISTER_flares}
may be very small compared to the true bolometric energy. 
However, since the SXR flare energy estimated from the empirical relation 
between the X-ray luminosity and H$\alpha$ luminosity (equation \ref{eq:LX2LHa})
may have a large uncertainty,
it is unclear whether the true bolometric energy for flare C is
much larger than the flare energy estimated from the TESS light curve.
Fig. \ref{fig:CMEmass} represents the mass and kinetic energy
of prominence eruptions and CMEs as a function of
X-ray flare energy in the GOES $1$--$8$ \AA band.
The estimated mass of the upward-moving material
for the flare C is roughly on the relation between the mass
and flare energy for stellar flares.
The masses for stellar events lie along or slightly below
the extrapolated solar trend.
The estimated kinetic energy for the flare C appears to
be good agreement with the relation between the
kinetic energy and X-ray flare energy for stellar flares.
However, the kinetic energies for stellar events, including the
flare C, are roughly two orders of magnitude smaller than 
the extrapolated solar trend as pointed out by \citet{Moschou2019}.
Since the kinetic energies for stellar events
are derived from the blue-shifted enhancement in
chromospheric lines (e.g., Balmer lines)
or X-ray dimming associated with the enhancement
of the hydrogen column density,
these stellar events may not be CMEs but prominence/filament 
eruptions.
As mentioned above, the average velocity of CMEs
on our Sun is $\sim 4$-$8$ times faster than that of prominence/filament eruptions (e.g., \cite{Gopalswamy2003}).
Therefore the kinetic energy for stellar events estimated from the
velocity of upward-moving prominence would be $1$-$2$ orders of magnitude
smaller than the solar CME trend.

\subsection{Rotational modulations}
\begin{figure}
 \begin{center}
  \includegraphics[width=16cm]{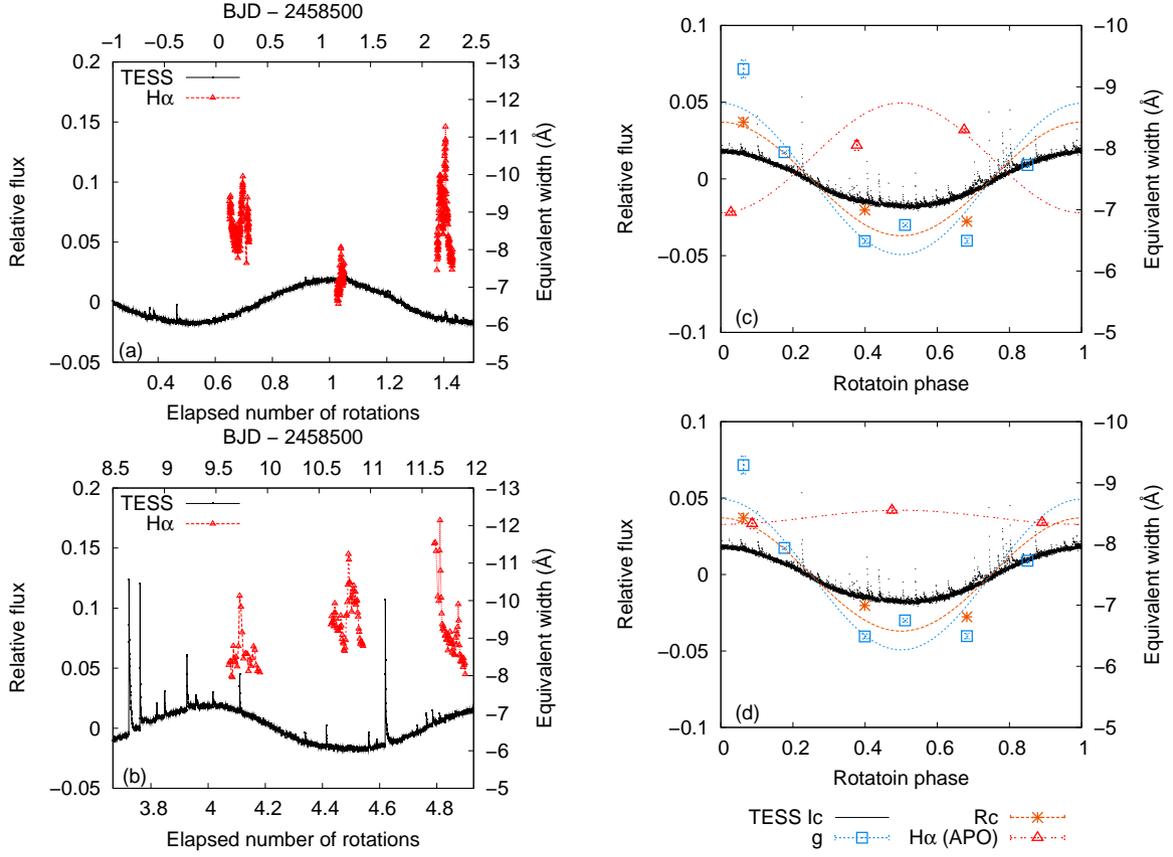} 
 \end{center}
\caption{
(a) TESS light curve (black line) and H$\alpha$ equivalent width
(red open triangles) of YZ CMi around OISTER observation period.
The left and right vertical axis indicate the relative flux in TESS
band and equivalent width of H$\alpha$ emission line. The horizontal
axis indicates the number of stellar rotation since BJD $2458498.329$.
(b) Same as (a) but for the light curve around APO observation period.
(c) Rotational brightness modulations in $g$- (blue open-squares), 
$R_C$- (orange asterisks), TESS bands (black dots; large flares were
excluded) observed during
the TESS cycle 7 and
the rotational modulation of the H$\alpha$ equivalent width (red triangles)
obtained by OISTER observations.
$g$-band, $R_C$-band, and H$\alpha$ data indicate the average value during
each observation run except for the data during flares.
The rotation phase was calculated from equation (\ref{rotation_phase}).
Blue dotted, orange dashed, and red dash-dotted lines indicate
sinusoidal fits to $g$-band, $R_C$-band, and H$\alpha$ line data.
(d) Same as (c) but for the rotational modulations of the 
equivalent width of H$\alpha$ emission line obtained by APO observations.
}\label{fig:Rotation1}
\end{figure}

As shown in Fig. \ref{fig:TESS_LC1}, YZ CMi shows the sinusoidal brightness
variations with the period of $2.7737\pm0.0014$ d and full-amplitude of $3.6$ \%.
The maximum timing of observed flux can be represented by the following equation:
\begin{equation}
    T_{\rm max; BJD} = 2458498.329 + 2.7737 \times E, \label{rotation_phase}
\end{equation}
where $E$ is the number of rotation.
These variations are thought to be caused by the rotation of the star with
spotted surface (e.g., \cite{Berdyugina2005}, \cite{Strassmeier2009}) 
and the amplitude of rotational modulations is thought
to correlate with the size of starspots.
Fig. \ref{fig:Rotation1} shows the rotational modulation of YZ CMi
in continuum light ($g$-, $R_C$-, and TESS bands) plotted with
that of the equivalent width of H$\alpha$ line.
The rotational modulations in $g$-, $R_C$-, and TESS bands can be
represented by sinusoidal functions in phase.

The amplitude of rotational modulations in continuum 
light increases as the wavelength decreases.
The sinusoidal fits to the data yield the full-amplitude of $9.8$ \% in $g$-band and $7.4$ \% in $R_C$-band, respectively.
Starspots are thought to be 
cooler than surrounding photosphere as the sunspots are.
The temperature difference between spots and unspotted photosphere
determine how the amplitude of rotational 
modulation depends on the wavelength.
In addition to the temperature difference between spots and unspotted
regions, the wavelength dependence of limb-darkening coefficient
also causes the amplitude of rotational modulation
depends on the wavelength.
Since the limb-darkening coefficient of low mass stars increases as the 
wavelength decreases in optical and near infra-red bands
(e.g., \cite{Claret2000}),
the effective size of the apparent stellar disk decreases as
the wavelength decreases.
Therefore the amplitude of rotational modulations increases
as the wavelength decreases.
We estimated the change in the spot coverage due to 
the stellar rotation from the amplitude of rotational modulations
in $g$-, $R_C$-, and TESS bands.
As mentioned above, not only the temperature difference between spots and 
unspotted photosphere, but also the limb-darkening coefficient
affect the amplitude of rotational modulations.
We estimated the temperature difference between spots and unspotted
photosphere  ($\Delta T = T_{\rm star} - T_{\rm spot}$) by using the following equation,
\begin{equation}
\Delta T(T_{\rm star}) = T_{\rm star} - T_{\rm spot}
 = 3.58\times 10^{-5}T_{\rm star}^2  + 0.249T_{\rm star} - 808,
\label{deltaT}
\end{equation}
which was derived from the second order polynomial fit to the data 
for all stars except for EK Dra in table 5 of \citet{Berdyugina2005}.
In the case of YZ CMi ($T_{\rm star} = 3300$ K; e.g., \cite{Gaidos2014}),
the the temperature difference between spots and photosphere
is estimated to be $\Delta T \sim 400$ K.
Accroding to \citet{Baroch2020}, the temperature difference between 
the photosphere and spot estimated is estimated to be
$\sim 200$ K ($178.6$-$273.4$; best model: $199.7\pm 9.6$ K)
from chromatic radial velocities.
We assumed that the limb-darkening law is represented by
\begin{equation}
    I(\mu)/I(1) = 1 - \sum _{k=1}^{4} a_k (1-\mu ^{k/2}),
\end{equation}
where $I(\mu)/I(1)$ is the intensity of the stellar disk 
relative to the intensity at the disk center and $a_k$ are the
limb-darkening coefficients \citep{Claret2000}. The parameter
$\mu$ is given by $\mu=\cos \theta$, where $\theta$ is the angular 
distance from the center of the disk.
$a_k$ are taken from tables in \citet{Claret2012} for the
star with the effective temperature of 3300 K and
surface gravity ($\log g$) of 5.0.
For YZ CMi, the effective area
of stellar disk would be reduced by
$\sim 22$\% in $g$-band, $\sim 20$\% in $R_C$-band and $\sim 14$\% in TESS-band
compared with the geometric (apparent) area.
The amplitude of the rotational brightness modulation
in $g$, $R_C$, and TESS bands ($9.8$ \%, $7.4$ \%, and $3.6$ \%)
correspond to the rotational change in spot coverage of $11$ \%,
$9.7$ \%, and $6.2$ \%
if we assume $\Delta T=400$ K.
For $\Delta T=200$ K, the rotational change in spot coverage 
is estimated to be $11$ \% (from TESS band) and $17$ \% (from 
$g$- and $R_C$- band).
This result suggests that at least $\sim 10$ -- $20$ \% of stellar surface
would be covered by starspots. The estimated spot coverage is comparable 
to those estimated by \citet{Zboril2003} and \citet{Baroch2020}.
Since the amplitude of
rotational modulations depends on the inclination angle and
spot latitude, it should be noted that these values are lower limits.
The inclination angle of YZ CMi is estimated to be
$60\,^\circ$ \citep{Morin2008} and $36^{+17}_{-14}\,^\circ$ \citep{Baroch2020}.
The Zeeman-Doppler Imaging observation by \citet{Morin2008} found
a strong spot near the visible pole. These results suggest that
the total spot coverage would be larger than the value estimated
from the amplitude of rotational modulations.

During the OISTER observations, the intensity of the H$\alpha$ line 
shows the anti-correlated rotational modulation with that in 
continuum light (Fig. \ref{fig:Rotation1} (a)).
Since the amplitude of rotational modulation in H$\alpha$ ($11$ \%)
is slightly larger than that of the rotational modulation in
$R_C$-band ($7.4$ \%),
the rotational modulation in H$\alpha$ line 
cannot be produced only by that in $R_C$-band
if we assume that the H$\alpha$ line flux is constant over the 
rotation phase and only the continuum intensity changes due to the 
rotation.
On the other hand, the rotational modulation in H$\alpha$ line
cannot be seen in the data obtained during the APO observations.
The H$\alpha$ line intensity is almost constant over the rotation 
phase (Fig. \ref{fig:Rotation1} (b)).
These results suggest that
bright regions in H$\alpha$ line are concentrated at the darker 
side in continuum (rotation phase of $0.5$) during the OISTER observation period,
and the H$\alpha$ line intensity at the brighter side in the continuum 
(rotation phase of $0$) has increased before or during the APO observations.
As shown in Fig. \ref{fig:TESS_LC1} (b) and (c), the number of flares
during the OISTER observations is smaller than that during the APO observations,
and large flares occurred just before and during the APO observations
at the rotation phase of $\sim 0.7$ -- $0.8$.
This suggests that the H$\alpha$ intensity of the existing active 
region, which can be seen around the rotation phase of $\sim 0.7$ -- $0.8$,
has increased due to the new flux emergence that makes the active region 
more flare-productive, or
new flare-productive active region has appeared
at the longitude corresponds to the rotation phase 
of $\sim 0.7$ -- $0.8$. 
Since such active region(s) would be
bright in H$\alpha$ and can change the rotational modulations
in the H$\alpha$ line. 
Since the rotational modulations in the continuum are almost the same
between these two observations while the rotational modulations in
H$\alpha$ line are different, the sizes and locations of starspots
may not change, or the size of starspots associated
with the new active region may not be so large.
According to \citet{Vida2016}, anti-correlated modulations in the
H$\alpha$ line and in the optical continuum were reported in an active
M dwarf V374 Peg during the observations in 2009.
However, this anti-correlation is unclear for the data obtained
in 2005 and 2006. These difference in rotational modulations between
the data in 2005-2006 and in 2009 may also be explained by the
difference in the surface distribution of active regions.

\begin{figure}
 \begin{center}
  \includegraphics[width=8cm]{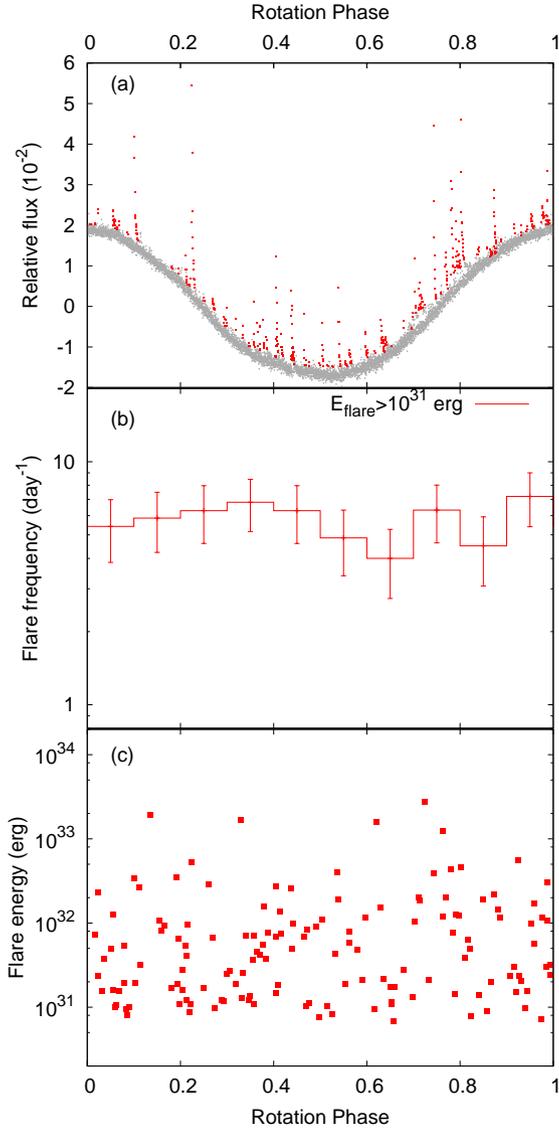} 
 \end{center}
\caption{
(a) Phase-folded light curve of rotational brightness variations
observed with TESS. The rotation phase was calculated from equation (\ref{rotation_phase}).
Red dots indicate the data points which are identified as flares.
(b) Frequency of flares with the bolometric energy larger than $10^{31}$ erg
as a function of rotation phase.
Error bars indicate the square-root of the number of flares in each phase bin.
(c) Bolometric energy of each flare as a function of rotation phase.
}\label{fig:Rotation2}
\end{figure}

Fig. \ref{fig:Rotation2} shows (a) rotational brighness variation,
(b) flare frequency as a function of rotation phase, and (c) bolometric
energy of each flare as a function of rotation phase.
There is no clear correlation between the rotation phase and
flare activities such as flare frequency and the energy of the largest
flare in each phase bin.
Similar results have been reported by \citet{Hawley2014} and 
\citet{Silverberg2016} in an active M dwarf GJ 1243,
and by \citet{Doyle2020} in active solar-type stars.
They found that the flare frequency and flare energy do not
depend on rotation phase.
\citet{Hawley2014} and \citet{Davenport2015} proposed two
possible interpretations for this property: 
(1) The large fraction of polar region is covered by starspots and flares come from the 
large starspots near the pole.
The rotational modulation of stellar brightness
in the optical continuum is mainly caused by the rotational change in the apparent area of
starspots near the pole
if the location of polar spot is slightly shifted from the rotation axis 
or the polar spot is not completely axisymmetric.
Since the starspots near the pole always can be seen from the observer, the flare frequency does not depend on the rotation phase.
(2) The most of flares come from many relatively small active regions which contain large amount of stellar magnetic flux \citep{Reiners2009}, 
while the polar spot is associated with the  global dipolar magnetic field \citep{Morin2008}. 
Since these many small active regions are distributed across longitude (e.g., \cite{Namekata2020a,Takasao2020}),
they cannot produce significant brightness modulation (e.g., \cite{Schrijver2020}) compared to that from the large polar spot 
and the flare frequency would not depend
on the rotation phase.
A strong correlation between the core emission of Ca H and K
lines and the magnetic field strength (e.g., \cite{Schrijver1989}, 
\cite{Notsu2015})
suggests that the origin of chromospheric emission is magnetic.
As mentioned above, although the rotational modulations in H$\alpha$ 
line are different between two observations, the rotational 
modulations in continuum are almost the same.
The possible explanations for the observed changes in rotational
modulations based on the possible interpretations for the rotational
modulations of flare frequency are as follows: 
During the low-activity period, both rotational modulations in the optical continuum
and in the H$\alpha$ line 
are mainly produced by the polar spot and surrounding active regions.
Since both the apparent area of polar spot and that of surrounding active regions
change as the star rotates, rotational modulations in the continuum and
in the H$\alpha$ line would be anti-correlated.
In the case of high-activity period,
although the rotational modulation in optical continuum is also 
produced by the polar spot, most of the
H$\alpha$ line emission come from 
many relatively small active regions which are distributed across longitude. 
Therefore the H$\alpha$ line intensity
does not depend on the rotation phase during the high-activity phase.
Since the chromospheric emission is thought to be correlated with the flare activity, 
these results on the rotational modulation
of H$\alpha$ line intensity suggest
that the constant flare frequency over the 
rotation phase on active M dwarfs
can be explained if the most of flares would
occur on many smaller active regions across 
the entire stellar longitude.

\subsection{Flare duration vs. flare energy}
\begin{figure}
 \begin{center}
  \includegraphics[width=8cm]{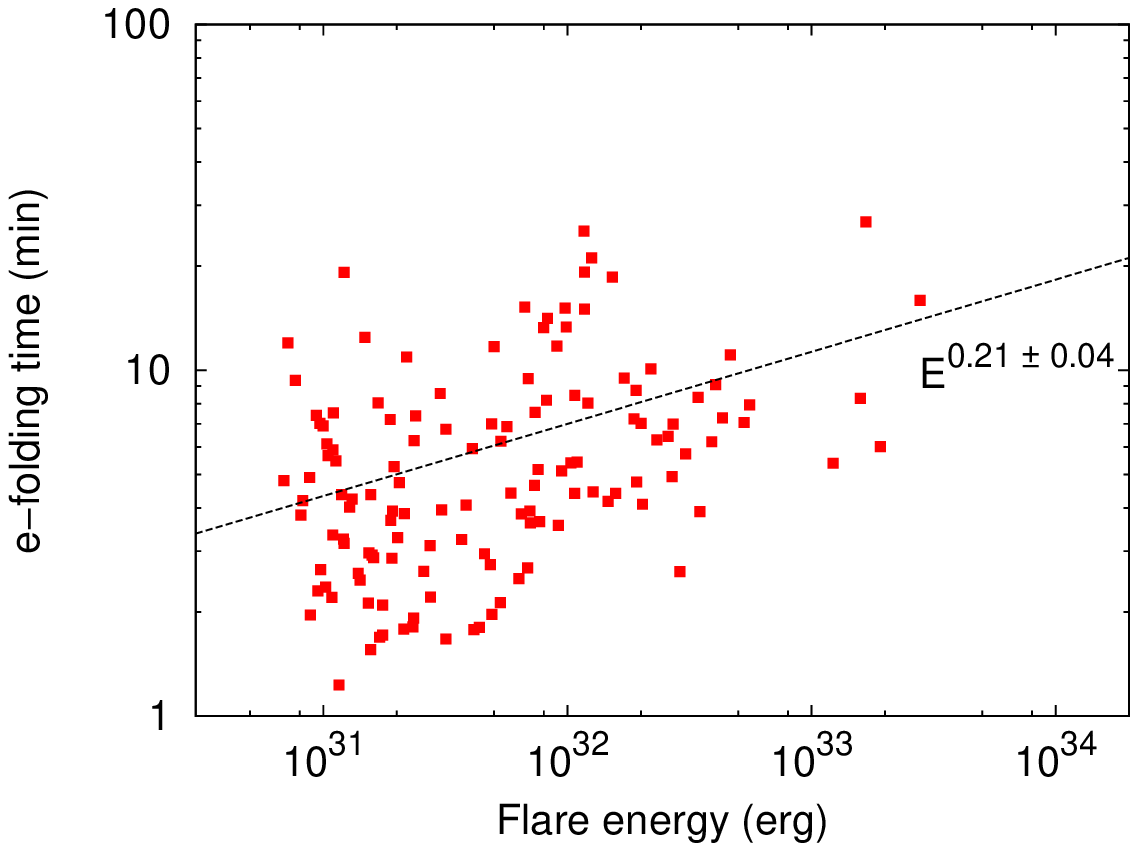} 
 \end{center}
\caption{
Scatter plot of flare duration (vertical axis) as a function of 
the bolometric flare energy (horizontal axis).
We used $e$-folding time from the flare peak time
as the duration of flares.
The bolometric flare energy was estimated from the TESS light curve
under the assumption that the spectral energy distribution of flare component
can be represented by the black body radiation with the temerature of $10^4$ K.
}\label{fig:flare_duration1}
\end{figure}

Fig. \ref{fig:flare_duration1} shows the scatter plot of the duration  ($e$-folding time)
of flares on YZ CMi observed with TESS as a function of flare energy.
We estimated the $e$-folding time by the exponential fit to the data
for each flare from the flare peak time ($t=t_{\rm peak}$) to 
$t=t_{\rm peak}+2(t_{1/e}-t_{\rm peak})$, where $t_{1/e}$ is 
the time that the flare flux decays to $1/e$ of the peak flux.
We can see a positive correlation between the flare duration ($\tau_{\rm flare}$)
and the bolometric flare energy ($E_{\rm flare}$).
The duration of flares increases with energy as
$\tau _{\rm flare} \propto E_{\rm flare}^{0.21\pm 0.04}$.
The power-law slope of the correlation between the flare duration
and energy for flares on YZ CMi is smaller than that for flares
on G-dwarfs observed with Kepler \citep{Maehara2015}
and with TESS \citep{Tu2020}.

\begin{figure}
 \begin{center}
  \includegraphics[width=8cm]{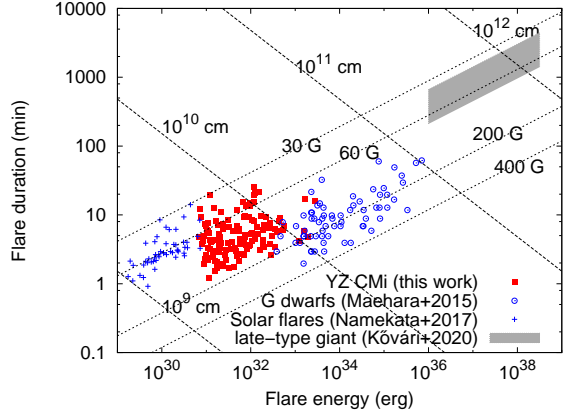} 
 \end{center}
\caption{
Comparison of $\tau$-$E$ relations for solar white-light 
flares (blue crosses; \cite{Namekata2017}),
stellar flares on an M dwarf YZ CMi (red filled-squares; this work),
superflares on solar-type stars (blue open-circles; \cite{Maehara2015}),
and superflares on a late-type giant  
KIC 2852961 (gray area; \cite{Kovari2020}).
Dashed lines indicate $\tau$-$E$ relation based on equation (\ref{eq: tau_E_B})
for $B=30$, $60$, $200$, and $400$ G.
Dotted lines represent another $\tau$-$E$ relation based on equation
(\ref{eq: tau_E_L}) for $L=10^{9}$, $10^{10}$, $10^{11}$, and $10^{12}$ cm.
}\label{fig:flare_duration1-2}
\end{figure}

Fig. \ref{fig:flare_duration1-2} shows the comparison of
$E$-$\tau$ relations for solar flares and stellar flares on various
types of stars.
Each $E$-$\tau$ relation shows positive correlation
and the power-law slope for these $E$-$\tau$ relations are roughly the same: $\tau _{\rm flare} \propto E^{1/3}_{\rm flare}$.
\citet{Maehara2015} proposed that,
if we assume the duration of white-light flare ($\tau _{\rm flare}$)
is determined by the reconnection time ($\tau _{\rm rec}$)
and the flare energy ($E_{\rm flare}$) correlates with the 
magnetic energy stored near the starspots ($E_{\rm mag} \sim B^2L^3/8\pi$),
the correlation between the flare duration and flare energy can be written
by the following form,
\begin{equation}
    \tau _{\rm flare} \sim \tau _{\rm rec} \propto \tau _{A}/M_{A} \sim L/v_{A}/M_{A} \propto E^{1/3}/B^{2/3}/v_{A}/M_{A},
\end{equation}
where $L$, $B$, $v_A$, and $M_A$ are the scale length of flaring
region, magnetic field strength, Alfv$\acute{\rm e}$n velocity,
and the dimensionless reconnection rate, which ranges from 0.01
to 0.1 in the case of the Petschek-type fast reconnection
\citep{ShibataMagara2011}.
This suggests that the power-law slope for the correlation 
between the duration
of flares and flare energy is $\sim 1/3$ if we assume that
$B$, $v_A$ and $M_A$ are roughly the same 
for flares on different stars.

However, we can see the discrepancy between the duration
of solar flares and those of flares on an M dwarf YZ CMi, G-dwarfs, 
and a giant KIC 2852961: the duration of solar flares and those
of flares on YZ CMi and KIC 2852961 are
roughly 3 times longer than that expected from the $\tau$-$E$ 
relation for superflares on G-dwarfs.
\citet{Chang2015} and \citet{Howard2019} found the
discrepancy between 
the $\tau$-$E$ relations for M dwarfs with different
temperature.
\citet{Namekata2017} pointed out 
the discrepancy between the $E$-$\tau$ relations for solar
white-light flares and superflares on G-dwarfs. 
They proposed the extended scaling-law, which takes into account 
the dependence on Alfv$\acute{\rm e}$n velocity ($v_A=B/\sqrt{4\pi\rho}$)
as follows:
\begin{equation}
    \tau _{\rm flare} \propto E^{1/3} B^{-5/3} \rho^{1/2}, \label{new_E-tau1}
\end{equation}
where $\rho$ is the pre-flare coronal density around the reconnection region.
By assuming the constant pre-flare coronal density,
equation (\ref{new_E-tau1}) can give the following new scaling-law:
\begin{equation}
    \tau _{\rm flare} \propto E^{1/3} B^{-5/3}. \label{eq: tau_E_B}
\end{equation}
Since $E\propto B^{2} L^{3}$ (e.g., \cite{Shibata2013}), the scaling-law can also be written as
\begin{equation}
    \tau _{\rm flare} \propto E^{-1/2} L^{5/2}. \label{eq: tau_E_L}
\end{equation}
As shown in Fig. \ref{fig:flare_duration1-2}, each $E$-$\tau$ relation
roughly is on the theoretical scaling-row with the constant $B$
(equation ($\ref{eq: tau_E_B}$)). This suggest that one possible explanation
for the discrepancy of $E$-$\tau$ relations for solar flares and stellar 
flares on various types of stars is the difference in the coronal
magnetic filed strength.
Moreover, another theoretical scaling-row given by the equation ($\ref{eq: tau_E_L}$) 
implies that the scale length of the flare region for the largest flare on YZ CMi is
comparable to 
the stellar radius ($R_{star} \sim 0.3R_{\odot} \sim 2\times 10^{10}$ cm for YZ CMi).
In the case of superlfares on solar-type stars and those on
a giant KIC 2852961 (gray),
the scale length for the largest flares is also comparable to the 
stellar radius ($\sim 1 R_{\odot}$ for solar-type stars and $\sim 10R_{\odot}$
for KIC 2852961). These upper limits suggest that the flare loop length
is roughly the same order of magnitude to the stellar radius.
The flare loop length estimated from the correlation between 
electron temperature and emission measure of solar and 
stellar flares \citep{ShibataYokoyama1999} is the same order of 
magnitude as the scale length estimated from $\tau$-$E$ relation
based on equation (\ref{eq: tau_E_L}).

\subsection{Flare duration statistics}
\begin{figure}
 \begin{center}
  \includegraphics[width=8cm]{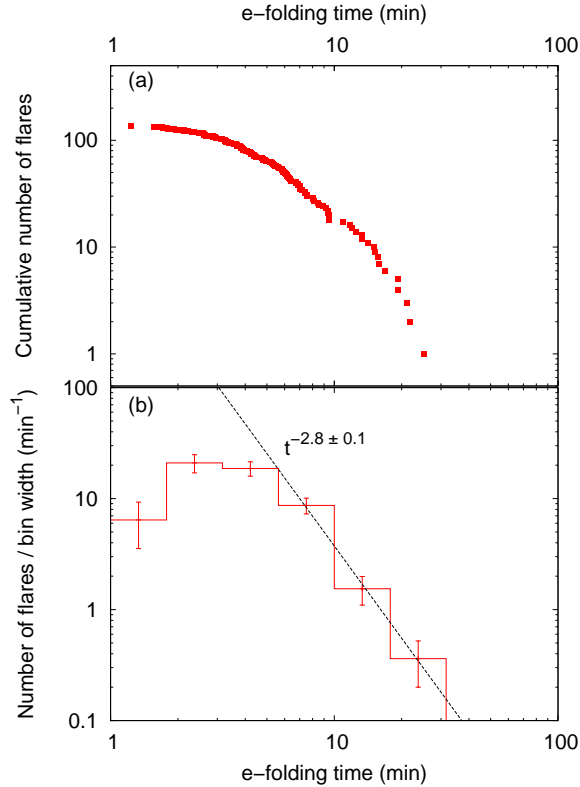} 
 \end{center}
\caption{
Number of flares as a function of flare duration.
The vertical and horizontal axis indicate the number of flares 
in each bin and e-folding time of flares.
}\label{fig:flare_duration2}
\end{figure}

 According to \citet{Veronig2002}, the frequency distribution
 of solar flares as a function of the flare duration can
 be represented by a power-law function with the power-law
 index of $-2.9$.
 The similar plots of the frequency
 distribution as a function of the flare duration ($e$-folding time)
 for the flares on YZ CMi are shown in Fig. \ref{fig:flare_duration2}. 
 We found that the frequency distribution of flares on YZ CMi
 as a function of the flare duration can also be well represented
 by a power-law function with the power-law index of $-2.8\pm 0.1$.
 This value is comparable to the power-law index for the 
 frequency distribution of solar flares.
 This similarity in flare frequency distribution as a function of
 flare duration may come from the similarity in the flare frequency
 distribution as a function of flare energy
 and that in the duration-energy correlations
 for solar flares observed in soft X-rays
 and for stellar flares observed in optical.
 The flare frequency distribution as a function of energy (flare fluence)
 can be well represented by a power-law function 
 ($dN/dE\propto E^{\alpha}$)
 with the power-law index of $\alpha = -2.03$ 
 for solar soft X-ray flares \citep{Veronig2002}.
 In the case of optical flares on YZ CMi, the flare frequency
 distribution as a function of bolometric energy of flares
 can also be represented by a power-law function with the power-law
 index of $-1.75$.
 Moreover, both the duration-energy (flare fluence) correlation
 ($\tau _{\rm flare} \propto E_{\rm flare} ^{\beta}$) for
  solar soft X-ray flares and that for optical flares on YZ CMi
  show similar power-law slope ($\sim 1/3$ for solar flares :\cite{Veronig2002}; $\sim 0.2$ for YZ CMi: this work).
 These similarities between the statistical properties of solar flares
 and those of stellar flares suggest 
 that the energy release mechanisms are
 the same for solar and stellar flares.
 
\section{Summary and Conclusions}
We carried out time-resolved photometry and 
spectroscopy of the M-type flare star YZ CMi
during the TESS observation window.
We detected 4 H$\alpha$ flares from our spectroscopic
and photometric observations and one of them exhibited
no clear brightening in continuum.
The key findings from our observations and 
are as follows:
\begin{enumerate}
    \item The H$\alpha$ flares with brightening in continuum
    tend to have shorter flare duration and larger H$\beta$ 
    emission line flux than the non-white-light flare.
    \item During the non-white-light flare, the blue asymmetry
    in the H$\alpha$ emission line was observed for $\sim 60$ min.
    The line of sight velocity of the blue asymmetry was $-80$ -- $-100$ km s$^{-1}$.
    \item If we assume the blue asymmetry in the H$\alpha$ line
    was caused by a prominence eruption, 
    the mass and kinetic energy of erupted material are estimated
    to be $10^{16}$ -- $10^{18}$ g and $10^{29.5}$ -- $10^{31.5}$ erg,
    respectively.
    Although these values are comparable to those expected from the empirical
    relations between
    flare X-ray energy and mass/kinetic energy for stellar events,
    the estimated kinetic energy is 2 orders of magnitude smaller than
    that expected from the empirical relation for CMEs on our Sun.
    This could be understood by the difference between
    the velocity of prominence eruption and that of CMEs.
    More simultaneous optical, and X-ray/EUV/radio observations 
    are necessary to fully understand stellar CMEs.
    \item The TESS light curve shows rotational modulations with
    the period of $2.7737\pm 0.0014$ d and full-amplitude of 3.6 \%.
    The rotational modulations can also be seen in $g$- and $R_C$-band
    data.
    The amplitude of the rotational modulations increases as the wavelength decreases ($9.8$ \% in $g$-band and $7.4$ \% in $R_C$-band).
    These rotational modulations suggest that $10$--$20$ \% of the surface of YZ CMi
    would be covered by starspots.
    \item The rotational modulation can also be seen in the equivalent width of H$\alpha$ line during the OISTER observation.
    The H$\alpha$ line modulation was anti-correlated with 
    rotational modulations in continuum. 
    However, the equivalent width of H$\alpha$ 
    line was almost constant during the APO observation.
    The TESS light curve suggests that the difference in rotational 
    modulations in H$\alpha$ line between these
    observations may be caused by the difference in the flare activity 
    during each observation run.
    \item The flare frequency and the largest flare energy
    during the TESS observation period were roughly constant
    over the rotation phase.
    \item The duration of flares ($\tau$) observed with TESS shows
    a positive correlation with the flare energy.
    The flare duration distribution shows the power-law
    distribution ($dN/d\tau \propto \tau^{\alpha}$) with
    the power-law index of $\alpha=-2.8\pm 0.1$, which is consistent
    with the power-law index of the flare duration distribution for
    solar flares.
\end{enumerate}

\begin{ack}
This work was supported by the Optical and Near-infrared Astronomy
Inter-University Cooperation Program and the Grants-in-Aid of the Ministry
of Education.
This work was also supported by MEXT/JSPS KAKENHI Grant Number 
17K05400, 20K04032 (H.M.), 18J20048 (K.N.), and 17K14253 (M.Y.).
Y.N. was supported by JSPS Overseas Research Fellowship Program. Y.N. also acknowledges the International Space Science Institute and the supported International Team 464: ``The Role Of Solar And Stellar Energetic Particles On (Exo)Planetary Habitability (ETERNAL, {http://www.issibern.ch/teams/exoeternal/})".
This paper includes data collected with the TESS mission, obtained from the MAST data archive at the Space Telescope Science Institute (STScI). 
Funding for the TESS mission is provided by the NASA Explorer Program. 
STScI is operated by the Association of Universities for Research in Astronomy, Inc., under NASA contract NAS 5–26555.
In this study, we used the observation data obtained with the Apache Point Observatory (APO) 3.5 m telescope, which is owned and operated by the Astrophysical Research Consortium.  We used APO observation time allocated to the University of Washington, and appreciate Suzanne Hawley for helping to plan the observation. We are grateful to APO 3.5m Observing Specialists and other staff members of Apache Point Observatory for their contributions in carrying out our observations. 
We are grateful to Mr. Isaiah Tristan (University of Colorado Boulder) 
for valuable suggestions to improve the manuscript.
We thank Dr. Jun Takahashi (Nishi-Harima Astronomical Observatory, Center for Astronomy, University of Hyogo) for his support for our observation by using Nayuta Telescope.
We also thank Prof. Jeffrey Linsky for useful comments and suggestions that contributed to improve the manuscript. 
\end{ack}

%



\begin{thebibliography}{}

\bibitem[Airapetian et al.(2016)]{Airapetian2016} Airapetian, V.~S., Glocer, A., Gronoff, G., et al.\ 2016, Nature Geoscience, 9, 452

\bibitem[Airapetian et al.(2020)]{Airapetian2020} Airapetian, V.~S., Barnes, R., Cohen, O., et al.\ 2020, International Journal of Astrobiology, 19, 136

\bibitem[Allred et al.(2015)]{Allred2015} Allred, J.~C., Kowalski, A.~F., \& Carlsson, M.\ 2015, \apj, 809, 104

\bibitem[Alvarado-G{\'o}mez et al.(2018)]{Alvarado-Gomez2018} Alvarado-G{\'o}mez, J.~D., Drake, J.~J., Cohen, O., et al.\ 2018, \apj, 862, 93

\bibitem[Alvarado-G{\'o}mez et al.(2020)]{Alvarado-Gomez2020} Alvarado-G{\'o}mez, J.~D., Drake, J.~J., Fraschetti, F., et al.\ 2020, \apj, 895, 47

\bibitem[Argiroffi et al.(2019)]{Argiroffi2019} Argiroffi, C., Reale, F., Drake, J.~J., et al.\ 2019, Nature Astronomy, 3, 742

\bibitem[Aschwanden et al.(2000)]{Aschwanden2000} Aschwanden, M.~J., Tarbell, T.~D., Nightingale, R.~W., et al.\ 2000, \apj, 535, 1047

\bibitem[Aschwanden et al.(2017)]{Aschwanden2017} Aschwanden, M.~J., Caspi, A., Cohen, C.~M.~S., et al.\ 2017, \apj, 836, 17


\bibitem[Audard et al.(2000)]{Audard2000} Audard, M., G{\"u}del, M., Drake, J.~J., et al.\ 2000, \apj, 541, 396

\bibitem[Baroch et al.(2020)]{Baroch2020} Baroch, D., Morales, J.~C., Ribas, I., et al.\ 2020, \aap, 641, A69


\bibitem[Benz \& G{\"u}del(2010)]{Benz2010} Benz, A.~O., \& G{\"u}del, M.\ 2010, \araa, 48, 241

\bibitem[Berdyugina(2005)]{Berdyugina2005} Berdyugina, S.~V.\ 2005, Living Reviews in Solar Physics, 2, 8

\bibitem[Berlicki(2007)]{Berlicki2007} Berlicki, A.\ 2007, The Physics of Chromospheric Plasmas, 387

\bibitem[Bessell(2005)]{Bessell2005} Bessell, M.~S.\ 2005, \araa, 43, 293

\bibitem[Canfield et al.(1990)]{Canfield1990} Canfield, R.~C., Penn, M.~J., Wulser, J.-P., et al.\ 1990, \apj, 363, 318

\bibitem[Canfield et al.(1996)]{Canfield1996} Canfield, R.~C., Reardon, K.~P., Leka, K.~D., et al.\ 1996, \apj, 464, 1016
\bibitem[Carlsson \& Stein(1997)]{Carlsson1997} Carlsson, M. \& Stein, R.~F.\ 1997, \apj, 481, 500
\bibitem[Chang et al.(2015)]{Chang2015} Chang, S.-W., Byun, Y.-I., \& Hartman, J.~D.\ 2015, \apj, 814, 35

\bibitem[Claret(2000)]{Claret2000} Claret, A.\ 2000, \aap, 363, 1081
\bibitem[Claret et al.(2012)]{Claret2012} Claret, A., Hauschildt, P.~H., \& Witte, S.\ 2012, \aap, 546, A14

\bibitem[Collier Cameron \& Robinson(1989)]{CollierCameron1989} Collier Cameron, A., \& Robinson, R.~D.\ 1989, \mnras, 236, 57

\bibitem[Collins et al.(2017)]{Collins2017} Collins, K.~A., Kielkopf, J.~F., Stassun, K.~G., et al.\ 2017, \aj, 153, 77

\bibitem[Cranmer(2017)]{Cranmer2017} Cranmer, S.~R.\ 2017, \apj, 840, 114

\bibitem[Crosby et al.(1993)]{Crosby1993} Crosby, N.~B., Aschwanden, M.~J., \& Dennis, B.~R.\ 1993, \solphys, 143, 275

\bibitem[Crosley \& Osten(2018)]{Crosley2018} Crosley, M.~K. \& Osten, R.~A.\ 2018, \apj, 856, 39

\bibitem[Davenport et al.(2015)]{Davenport2015} Davenport, J.~R.~A., Hebb, L., \& Hawley, S.~L.\ 2015, \apj, 806, 212

\bibitem[Dennis \& Zarro(1993)]{Dennis1993} Dennis, B.~R. \& Zarro, D.~M.\ 1993, \solphys, 146, 177

\bibitem[Doyle et al.(2020)]{Doyle2020} Doyle, L., Ramsay, G., \& Doyle, J.~G.\ 2020, \mnras, 494, 3596


\bibitem[Drake \& Ulrich(1980)]{Drake1980} Drake, S.~A., \& Ulrich, R.~K.\ 1980, \apjs, 42, 351


\bibitem[Drake et al.(2016)]{Drake2016} Drake, J.~J., Cohen, O., Garraffo, C., et al.\ 2016, Solar and Stellar Flares and Their Effects on Planets, 196

\bibitem[Dunstone et al.(2006)]{Dunstone2006} Dunstone, N.~J., Collier Cameron, A., Barnes, J.~R., et al.\ 2006, \mnras, 373, 1308

\bibitem[Emslie et al.(2012)]{Emslie2012} Emslie, A.~G., Dennis, B.~R., Shih, A.~Y., et al.\ 2012, \apj, 759, 71


\bibitem[Fausnaugh et al.(2019)]{Fausnaugh2019} Fausnaugh, M.~M., Burke, C.~J., Caldwell, D.~A., et al.\ 2019, TESS Data Release Notes: Sector 7, DR9



\bibitem[Fisher et al.(1985)]{Fisher1985} Fisher, G.~H., Canfield, R.~C., \& McClymont, A.~N.\ 1985, \apj, 289, 414

\bibitem[Fisher(1989)]{Fisher1989} Fisher, G.~H.\ 1989, \apj, 346, 1019

\bibitem[Fuhrmeister et al.(2008)]{Fuhrmeister2008} Fuhrmeister, B., Liefke, C., Schmitt, J.~H.~M.~M., et al.\ 2008, \aap, 487, 293

\bibitem[Fuhrmeister et al.(2018)]{Fuhrmeister2018} Fuhrmeister, B., Czesla, S., Schmitt, J.~H.~M.~M., et al.\ 2018, \aap, 615, A14

\bibitem[Fukugita et al.(1996)]{Fukugita1996} Fukugita, M., Ichikawa, T., Gunn, J.~E., et al.\ 1996, \aj, 111, 1748

\bibitem[Gaia Collaboration et al.(2018)]{Gaia2018} Gaia Collaboration, Brown, A.~G.~A., Vallenari, A., et al.\ 2018, \aap, 616, A1

\bibitem[Gaidos \& Mann(2014)]{Gaidos2014} Gaidos, E., \& Mann, A.~W.\ 2014, \apj, 791, 54

\bibitem[Gershberg(1972)]{Gershberg1972} Gershberg, R.~E.\ 1972, \apss, 19, 75

\bibitem[Gershberg(2005)]{Gershberg2005} Gershberg, R.~E.\ 2005, Solar-Type Activity in Main-Sequence Stars (Spromger: Berlin)

\bibitem[Gilbert et al.(2005)]{Gilbert2005} Gilbert, H.~R., Holzer, T.~E., \& MacQueen, R.~M.\ 2005, \apj, 618, 524

\bibitem[Gopalswamy et al.(2003)]{Gopalswamy2003} Gopalswamy, N., Shimojo, M., Lu, W., et al.\ 2003, \apj, 586, 562

\bibitem[Gopalswamy(2015)]{Gopalswamy2015} Gopalswamy, N.\ 2015, Solar Prominences, 381

\bibitem[Graham et al.(2020)]{Graham2020} Graham, D.~R., Cauzzi, G., Zangrilli, L., et al.\ 2020, \apj, 895, 6

\bibitem[Grindlay \& Heise(1975)]{Grindlay1975} Grindlay, J., \& Heise, J.\ 1975, International Cosmic Ray Conference, 154


\bibitem[Guedel et al.(1996)]{Guedel1996} Guedel, M., Benz, A.~O., Schmitt, J.~H.~M.~M., et al.\ 1996, \apj, 471, 1002

\bibitem[Gunn et al.(1994)]{Gunn1994} Gunn, A.~G., Doyle, J.~G., Mathioudakis, M., et al.\ 1994, \aap, 285, 489

\bibitem[Hawley \& Pettersen(1991)]{Hawley1991} Hawley, S.~L. \& Pettersen, B.~R.\ 1991, \apj, 378, 725
\bibitem[Hawley \& Fisher(1992)]{Hawley1992} Hawley, S.~L., \& Fisher, G.~H.\ 1992, \apjs, 78, 565
\bibitem[Hawley et al.(1995)]{Hawley1995} Hawley, S.~L., Fisher, G.~H., Simon, T., et al.\ 1995, \apj, 453, 464

\bibitem[Hawley et al.(2003)]{Hawley2003} Hawley, S.~L., Allred, J.~C., Johns-Krull, C.~M., et al.\ 2003, \apj, 597, 535

\bibitem[Hawley et al.(2014)]{Hawley2014} Hawley, S.~L., Davenport, J.~R.~A., Kowalski, A.~F., et al.\ 2014, \apj, 797, 121


\bibitem[Heinzel et al.(1994a)]{Heinzel1994} Heinzel, P., Gouttebroze, P., \& Vial, J.-C.\ 1994, \aap, 292, 656

\bibitem[Heinzel et al.(1994b)]{Heinzel1994_BlueAsym} Heinzel, P., Karlicky, M., Kotrc, P., et al.\ 1994, \solphys, 152, 393

\bibitem[Heinzel et al.(2008)]{Heinzel2008} Heinzel, P., Schmieder, B., F{\'a}rn{\'\i}k, F., et al.\ 2008, \apj, 686, 1383

\bibitem[Hilton(2011)]{Hilton2011} Hilton, E.~J.\ 2011, Ph.D. Thesis

\bibitem[Houdebine et al.(1990)]{Houdebine1990} Houdebine, E.~R., Foing, B.~H., \& Rodono, M.\ 1990, \aap, 238, 249

\bibitem[Houdebine et al.(1993)]{Houdebine1993} Houdebine, E.~R., Foing, B.~H., Doyle, J.~G., et al.\ 1993, \aap, 274, 245

\bibitem[Houdebine et al.(2016)]{Houdebine2016} Houdebine, E.~R., Mullan, D.~J., Paletou, F., et al.\ 2016, \apj, 822, 97


\bibitem[Honda et al.(2018)]{Honda2018} Honda, S., Notsu, Y., Namekata, K., et al.\ 2018, \pasj, 70, 62


\bibitem[Howard et al.(2019)]{Howard2019} Howard, W.~S., Corbett, H., Law, N.~M., et al.\ 2019, \apj, 881, 9


\bibitem[Huang et al.(2019)]{Huang2019} Huang, N., Xu, Y., Sadykov, V.~M., et al.\ 2019, \apjl, 878, L15


\bibitem[Ichimoto \& Kurokawa(1984)]{Ichimoto1984} Ichimoto, K., \& Kurokawa, H.\ 1984, \solphys, 93, 105


\bibitem[Jin et al.(2020)]{Jin2020} Jin, M., Cheung, M.~C.~M., DeRosa, M.~L., et al.\ 2020, arXiv e-prints, arXiv:2002.06249


\bibitem[Johns-Krull \& Valenti(2000)]{Johns-Krull2000} Johns-Krull, C.~M. \& Valenti, J.~A.\ 2000, Stellar Clusters and Associations: Convection, Rotation, and Dynamos, 198, 371



\bibitem[Kawabata et al.(2008)]{Kawabata2008} Kawabata, K.~S., Nagae, O., Chiyonobu, S., et al.\ 2008, \procspie, 70144L

\bibitem[Kerr \& Fletcher(2014)]{Kerr2014} Kerr, G.~S., \& Fletcher, L.\ 2014, \apj, 783, 98

\bibitem[Kleint et al.(2016)]{Kleint2016} Kleint, L., Heinzel, P., Judge, P., et al.\ 2016, \apj, 816, 88

\bibitem[Kotani et al.(2005)]{Kotani2005} Kotani, T., Kawai, N., Yanagisawa, K., et al.\ 2005, Nuovo Cimento C Geophysics Space Physics C, 28, 755

\bibitem[Kowalski et al.(2010)]{Kowalski2010} Kowalski, A.~F., Hawley, S.~L., Holtzman, J.~A., et al.\ 2010, \apjl, 714, L98

\bibitem[Kowalski et al.(2013)]{Kowalski2013} Kowalski, A.~F., Hawley, S.~L., Wisniewski, J.~P., et al.\ 2013, \apjs, 207, 15

\bibitem[Kowalski et al.(2017)]{Kowalski2017} Kowalski, A.~F., Allred, J.~C., Daw, A., et al.\ 2017, \apj, 836, 12

\bibitem[Kowalski et al.(2019)]{Kowalski2019} Kowalski, A.~F., Wisniewski, J.~P., Hawley, S.~L., et al.\ 2019, \apj, 871, 167

\bibitem[Kuridze et al.(2015)]{Kuridze2015} Kuridze, D., Mathioudakis, M., Sim{\~o}es, P.~J.~A., et al.\ 2015, \apj, 813, 125

\bibitem[Kuridze et al.(2016)]{Kuridze2016} Kuridze, D., Mathioudakis, M., Christian, D.~J., et al.\ 2016, \apj, 832, 147

\bibitem[Lacy et al.(1976)]{Lacy1976} Lacy, C.~H., Moffett, T.~J., \& Evans, D.~S.\ 1976, \apjs, 30, 85


\bibitem[Lammer et al.(2007)]{Lammer2007} Lammer, H., Lichtenegger, H.~I.~M., Kulikov, Y.~N., et al.\ 2007, Astrobiology, 7, 185


\bibitem[Landi \& Reale(2013)]{Landi2013} Landi, E., \& Reale, F.\ 2013, \apj, 772, 71



\bibitem[Li et al.(2019)]{Li2019} Li, Y., Ding, M.~D., Hong, J., et al.\ 2019, \apj, 879, 30

\bibitem[Linsky(2019)]{Linsky2019} Linsky, J.\ 2019, Lecture Notes in Physics, Berlin Springer Verlag

\bibitem[Lovell(1969)]{Lovell1969} Lovell, B.\ 1969, \nat, 222, 1126

\bibitem[K{\H{o}}v{\'a}ri et al.(2020)]{Kovari2020} K{\H{o}}v{\'a}ri, Z., Ol{\'a}h, K., G{\"u}nther, M.~N., et al.\ 2020, \aap, 641, A83

\bibitem[Moschou et al.(2019)]{Moschou2019} Moschou, S.-P., Drake, J.~J., Cohen, O., et al.\ 2019, \apj, 877, 105

\bibitem[Maehara et al.(2012)]{Maehara2012} Maehara, H., Shibayama, T., Notsu, S., et al.\ 2012, \nat, 485, 478

\bibitem[Maehara et al.(2015)]{Maehara2015} Maehara, H., Shibayama, T., Notsu, Y., et al.\ 2015, Earth, Planets, and Space, 67, 59



\bibitem[Mitra-Kraev et al.(2005)]{Mitra-Kraev2005} Mitra-Kraev, U., Harra, L.~K., G{\"u}del, M., et al.\ 2005, \aap, 431, 679


\bibitem[Morin et al.(2008)]{Morin2008} Morin, J., Donati, J.-F., Petit, P., et al.\ 2008, \mnras, 390, 567

\bibitem[Muheki et al.(2020)]{Muheki2020} Muheki, P., Guenther, E.~W., Mutabazi, T., et al.\ 2020, \aap, 637, A13

\bibitem[Mullan \& Paudel(2019)]{Mullan2019} Mullan, D.~J. \& Paudel, R.~R.\ 2019, \apj, 873, 1

\bibitem[Namekata et al.(2017)]{Namekata2017} Namekata, K., Sakaue, T., Watanabe, K., et al.\ 2017, \apj, 851, 91

\bibitem[Namekata et al.(2020a)]{Namekata2020a} Namekata, K., Davenport, J.~R.~A., Morris, B.~M., et al.\ 2020, \apj, 891, 103

\bibitem[Namekata et al.(2020b)]{Namekata2020} Namekata, K., Maehara, H., Sasaki, R., et al.\ 2020, \pasj, doi:10.1093/pasj/psaa051

\bibitem[Neupert(1968)]{Neupert1968} Neupert, W.~M.\ 1968, \apjl, 153, L59

\bibitem[Newton et al.(2015)]{Newton2015} Newton, E.~R., Charbonneau, D., Irwin, J., et al.\ 2015, \apj, 800, 85

\bibitem[Notsu et al.(2015)]{Notsu2015} Notsu, Y., Honda, S., Maehara, H., et al.\ 2015, \pasj, 67, 33

\bibitem[Notsu et al.(2019)]{Notsu2019} Notsu, Y., Maehara, H., Honda, S., et al.\ 2019, \apj, 876, 58

\bibitem[Odert et al.(2017)]{Odert2017} Odert, P., Leitzinger, M., Hanslmeier, A., et al.\ 2017, \mnras, 472, 876

\bibitem[Odert et al.(2020)]{Odert2020} Odert, P., Leitzinger, M., Guenther, E.~W., et al.\ 2020, \mnras, 494, 3766

\bibitem[Osten \& Wolk(2015)]{Osten2015} Osten, R.~A., \& Wolk, S.~J.\ 2015, \apj, 809, 79

\bibitem[Ozaki \& Tokimasa(2005)]{Ozaki2005} Ozaki, S., \& Tokimasa, N.\ 2005, Ann. Rep. Nishi-Harima Astron. Obs., 15, 15

\bibitem[Pettersen et al.(1983)]{Pettersen1983}
Pettersen, B.~R., Kern, G.~A., \& Evans, D.~S.\ 1983, \aap, 123, 184

\bibitem[Reid \& Hawley(2005)]{Reid2005} Reid, I.~N., \& Hawley, S.~L.\ 2005, New Light on Dark Stars Red Dwarfs (Chichester: Praxis Publishing)

\bibitem[Reiners \& Basri(2007)]{Reiners2007} Reiners, A. \& Basri, G.\ 2007, \apj, 656, 1121

\bibitem[Reiners \& Basri(2009)]{Reiners2009} Reiners, A. \& Basri, G.\ 2009, \aap, 496, 787

\bibitem[Ricker et al.(2015)]{Ricker2015} Ricker, G.~R., Winn, J.~N., Vanderspek, R., et al.\ 2015, Journal of Astronomical Telescopes, Instruments, and Systems, 1, 014003

\bibitem[Segura et al.(2010)]{Segura2010} Segura, A., Walkowicz, L.~M., Meadows, V., et al.\ 2010, Astrobiology, 10, 751


\bibitem[Shakhovskaia(1989)]{Shakhovskaia1989} Shakhovskaia, N.~I.\ 1989, \solphys, 121, 375

\bibitem[Schmieder et al.(1987)]{Schmieder1987} Schmieder, B., Forbes, T.~G., Malherbe, J.~M., et al.\ 1987, \apj, 317, 956

\bibitem[Schrijver et al.(1989)]{Schrijver1989} Schrijver, C.~J., Cote, J., Zwaan, C., et al.\ 1989, \apj, 337, 964

\bibitem[Schrijver(2020)]{Schrijver2020} Schrijver, C.~J.\ 2020, \apj, 890, 121

\bibitem[Scheucher et al.(2018)]{Scheucher2018} Scheucher, M., Grenfell, J.~L., Wunderlich, F., et al.\ 2018, \apj, 863, 6


\bibitem[Shibata \& Yokoyama(1999)]{ShibataYokoyama1999} Shibata, K., \& Yokoyama, T.\ 1999, \apjl, 526, L49

\bibitem[Shibata \& Yokoyama(2002)]{ShibataYokoyama2002} Shibata, K., \& Yokoyama, T.\ 2002, \apj, 577, 422

\bibitem[Shibata \& Magara(2011)]{ShibataMagara2011} Shibata, K., \& Magara, T.\ 2011, Living Reviews in Solar Physics, 8, 6

\bibitem[Shibata et al.(2013)]{Shibata2013} Shibata, K., Isobe, H., Hillier, A., et al.\ 2013, \pasj, 65, 49


\bibitem[Shibayama et al.(2013)]{Shibayama2013} Shibayama, T., Maehara, H., Notsu, S., et al.\ 2013, \apjs, 209, 5

\bibitem[Shoji \& Kurokawa(1995)]{Shoji1995} Shoji, M., \& Kurokawa, H.\ 1995, \pasj, 47, 239


\bibitem[Silverberg et al.(2016)]{Silverberg2016} Silverberg, S.~M., Kowalski, A.~F., Davenport, J.~R.~A., et al.\ 2016, \apj, 829, 129

\bibitem[Soubiran et al.(2018)]{Soubiran2018} Soubiran, C., Jasniewicz, G., Chemin, L., et al.\ 2018, \aap, 616, A7


\bibitem[Spangler et al.(1974)]{Spangler1974} Spangler, S.~R., Shawhan, S.~D., \& Rankin, J.~M.\ 1974, \apjl, 190, L129


\bibitem[Stellingwerf (1978)]{Stellingwerf1978}
        Stellingwerf, R.~F.\ 1978, \apj, 224, 953

\bibitem[Strassmeier(2009)]{Strassmeier2009} Strassmeier, K.~G.\ 2009, \aapr, 17, 251

\bibitem[{\v{S}}vestka et al.(1962)]{Svestka1962} {\v{S}}vestka, Z., Kopeck{\'y}, M., \& Blaha, M.\ 1962, Bulletin of the Astronomical Institutes of Czechoslovakia, 13, 37

\bibitem[Takasao et al.(2020)]{Takasao2020} Takasao, S., Mitsuishi, I., Shimura, T., et al.\ 2020, arXiv:2008.04255

\bibitem[Tei et al.(2018)]{Tei2018} Tei, A., Sakaue, T., Okamoto, T.~J., et al.\ 2018, \pasj, 70, 100

\bibitem[Tilley et al.(2019)]{Tilley2019} Tilley, M.~A., Segura, A., Meadows, V., et al.\ 2019, Astrobiology, 19, 64



\bibitem[Tu et al.(2020)]{Tu2020} Tu, Z.-L., Yang, M., Zhang, Z.~J., et al.\ 2020, \apj, 890, 46

\bibitem[van Maanen(1945)]{vanMaanen1945} van Maanen, A.\ 1945, \pasp, 57, 216

\bibitem[Veronig et al.(2002)]{Veronig2002} Veronig, A., Temmer, M., Hanslmeier, A., et al.\ 2002, \aap, 382, 1070

\bibitem[Vida et al.(2016)]{Vida2016} Vida, K., Kriskovics, L., Ol{\'a}h, K., et al.\ 2016, \aap, 590, A11

\bibitem[Vida et al.(2019)]{Vida2019} Vida, K., Leitzinger, M., Kriskovics, L., et al.\ 2019, \aap, 623, A49

\bibitem[Vanderspek et al.(2018)]{Vanderspek2018} Vanderspek, R., Doty, J., Fausnaugh, M., et al.\ 2018, TESS Instrument Handbook, Tech. rep., Kavli Institute for Astrophysics and Space Science, Massachusetts Institute of Technology

\bibitem[Wang et al.(2003)]{Wang2003} Wang, S.-. i ., Hildebrand, R.~H., Hobbs, L.~M., et al.\ 2003, \procspie, 1145

\bibitem[Warren(2006)]{Warren2006} Warren, H.~P.\ 2006, \apj, 637, 522

\bibitem[Watanabe et al.(2013)]{Watanabe2013} Watanabe, K., Shimizu, T., Masuda, S., et al.\ 2013, \apj, 776, 123

\bibitem[Watanabe et al.(2017)]{Watanabe2017} Watanabe, K., Kitagawa, J., \& Masuda, S.\ 2017, \apj, 850, 204



\bibitem[Yamashiki et al.(2019)]{Yamashiki2019} Yamashiki, Y.~A., Maehara, H., Airapetian, V., et al.\ 2019, \apj, 881, 114

\bibitem[Yashiro \& Gopalswamy(2009)]{Yashiro2009} Yashiro, S., \& Gopalswamy, N.\ 2009, Universal Heliophysical Processes, 233

\bibitem[Zboril(2003)]{Zboril2003} Zboril, M.\ 2003, Astronomische Nachrichten, 324, 527

\end{thebibliography}
\end{document}